\documentclass[reprint,aps,pra,superscriptaddress,longbibliography]{revtex4-2}
\usepackage[latin9]{inputenc}
\usepackage{color}
\usepackage{amsmath}
\usepackage{amssymb}
\usepackage{stmaryrd}
\usepackage{graphicx}
\usepackage[unicode=true,
 bookmarks=true,bookmarksnumbered=false,bookmarksopen=false,
 breaklinks=false,pdfborder={0 0 1},backref=false,colorlinks=true]
 {hyperref}
\hypersetup{
 linkcolor=blue, urlcolor=blue, citecolor=blue, pdfstartview={FitH}, hyperfootnotes=true, unicode=true}

\usepackage[caption=false,subrefformat=parens,labelformat=parens]{subfig}
\definecolor{lightgray}{gray}{0.8}  
\usepackage{IEEEtrantools}
\usepackage{mathtools}
\usepackage{amsthm}
\theoremstyle{definition}\newtheorem{condition}{Condition} 
\theoremstyle{definition}\newtheorem{lemma}[condition]{Lemma}

\makeatletter

\@ifundefined{textcolor}{}
{%
 \definecolor{BLACK}{gray}{0}
 \definecolor{WHITE}{gray}{1}
 \definecolor{RED}{rgb}{1,0,0}
 \definecolor{GREEN}{rgb}{0,1,0}
 \definecolor{BLUE}{rgb}{0,0,1}
 \definecolor{CYAN}{cmyk}{1,0,0,0}
 \definecolor{MAGENTA}{cmyk}{0,1,0,0}
 \definecolor{YELLOW}{cmyk}{0,0,1,0}
}

\usepackage[normalem]{ulem}
\usepackage{cancel}

\usepackage{amsfonts}
\usepackage{tabularx}
\usepackage{dcolumn}
\usepackage{bm}
\usepackage{graphicx}
\usepackage{epstopdf}

\hypersetup{urlcolor=blue}

\usepackage{tensor}
\usepackage{braket}

\usepackage[capitalise,compress]{cleveref}
\crefname{section}{Sec.}{Secs.}
\Crefname{section}{Section}{Sections}
\crefrangelabelformat{equation}{\textup{(#3#1#4)}--\textup{(#5#2#6)}}

\makeatother

\begin{document}

\title{Efficiency of neural-network state representations of one-dimensional quantum spin systems}
\author{Ruizhi Pan}
\email{panruizhi@gmail.com}
\affiliation{Joint Quantum Institute, NIST/University of Maryland, College Park, MD 20742, USA}
\author{Charles W. Clark}
\email{charles.clark@nist.gov}
\affiliation{Joint Quantum Institute, NIST/University of Maryland, College Park, MD 20742, USA}
\affiliation{National Institute of Standards and Technology, Gaithersburg, Maryland 20899, USA}

\date{\today}

\begin{abstract}
Neural-network state representations of quantum many-body systems are attracting great attention and more rigorous quantitative analysis about their expressibility and complexity is warranted. Our analysis of the restricted Boltzmann machine (RBM) state representation of one-dimensional (1D) quantum spin systems provides new insight into their computational complexity. We define a class of long-range-fast-decay (LRFD) RBM states with quantifiable upper bounds on truncation errors and provide numerical evidence for a large class of 1D quantum systems that may be approximated by LRFD RBMs of at most polynomial complexities. These results lead us to conjecture that the ground states of a wide range of quantum systems may be exactly represented by LRFD RBMs or a variant of them, even in cases where other state representations become less efficient. At last, we provide the relations between multiple typical state manifolds. Our work proposes a paradigm for doing complexity analysis for generic long-range RBMs which naturally yields a further classification of this manifold. This paradigm and our characterization of their nonlocal structures may pave the way for understanding the natural measure of complexity for quantum many-body states described by RBMs and are generalizable for higher-dimensional systems and deep neural-network quantum states.

\end{abstract}

\maketitle


\section{Introduction} \label{sec:QML3intro}

Quantum machine learning is an emerging field that combines techniques in the disciplines of machine learning (ML) and quantum physics~\cite{RevModPhys.91.045002,biamonte2017quantum,carrasquilla2017machine,doi:10.1126/science.abk3333,sarma2019machine,carleo2017solving,melko2019restricted}. Research in this field takes three broad forms~\cite{biamonte2017quantum}: applications of classical ML techniques to quantum physical systems~\cite{doi:10.1126/science.abk3333,sarma2019machine,carleo2017solving,glasser2018neural,melko2019restricted,gao2017efficient,PhysRevB.94.165134}, quantum computing and algorithms for classical ML problems~\cite{lloyd2013quantum,10.5555/2871393.2871400,wiebe2014quantum,rebentrost2014quantum}, and new ideas inspired by the intersection of the two disciplines~\cite{amin2018quantum,mehta2014exact}. In the field of learning quantum systems, there has been tremendous progress in applying ML techniques to identifying quantum phases and transitions~\cite{carrasquilla2017machine,hu2017discovering,ch2017machine,wang2016discovering,PhysRevA.98.033604,hsu2018machine}, molecular modeling~\cite{rupp2012fast,bartok2017machine}, quantum state tomography~\cite{torlai2018neural,kieferova2017tomography}, and accelerating Monte Carlo simulations~\cite{PhysRevB.95.041101,PhysRevB.95.035105}.


Here, we report results of an investigation of neural network quantum states in the context of quantum many-body physics~\cite{sarma2019machine,carleo2017solving,cai2018approximating,PhysRevB.96.205152,gao2017efficient,deng2017machine,deng2017quantum,glasser2018neural,melko2019restricted}, a subject of much recent interest. The core idea is to postulate an ansatz for the wave function in terms of a neural network (NN)~\cite{carleo2017solving}, which targets a low-dimensional manifold in the exponentially large Hilbert space for state approximation~\cite{RevModPhys.93.045003}, and apply ML algorithms to find a specific solution. The restricted Boltzmann machine (RBM)~\cite{carleo2017solving,melko2019restricted,cai2018approximating,PhysRevB.96.205152,glasser2018neural} is a bipartite stochastic construct that combines the concepts of thermodynamic partition functions with those of classical artificial neural networks. RBMs have successfully represented a wide range of quantum states, such as low-lying eigenstates of quantum many-body-localized systems~\cite{carleo2017solving,choo2018symmetries}, code words of a stabilizer code~\cite{deng2017machine,zheng2019restricted,he2019multi} and chiral topological states~\cite{glasser2018neural,kaubruegger2018chiral,PhysRevLett.127.170601}.

While RBMs have demonstrated their power in numerical simulation, we have particular motivations to investigate the expressibility and complexity of the generic long-range RBMs, which are characterized by dense network architectures with full interlayer connectivity, in contrast to the so-called short-range or sparse RBMs~\cite{deng2017machine,deng2017quantum,melko2019restricted}. First, the state approximators that are produced by RBM solutions 
returned by ML algorithms often feature a long-range form and a fast parameter decay, even when the exact RBM representations of the target states are unknown~\cite{deng2017machine,deng2017quantum} or less efficient~\cite{glasser2018neural}. Increasing the number of hidden nodes captures spin correlations of higher orders~\cite{carleo2017solving}, increasing the approximation accuracy. The best approximators often have a form similar to that obtained by magnitude-based pruning~\cite{PhysRevB.105.125124,PhysRevB.100.195125} of a finite truncated RBM with infinitely many hidden nodes. These observations motivate the generalization of the RBM wave-function ansatz to an
infinitely-many-hidden-node regime and the justification of the faithfulness of using these truncated long-range RBM approximators~\cite{PhysRevB.73.094423}.

Another motivation for studying long-range RBMs stems from the central goal of exploring effective compressed state representations, which includes understanding the natural measure of complexity~\cite{melko2019restricted} and how the global physical information is encoded in that description ~\cite{RevModPhys.93.045003}. There has been some work studying the relationship between RBMs and other concepts about state representations, such as string-bond states~\cite{glasser2018neural}, correlator product states~\cite{Clark_2018} and tensor network states~\cite{PhysRevLett.127.170601,PhysRevB.97.085104}. However, analysis of the effects of truncations through transforming RBMs into other representations may lead to redundancy and inconvenience, and it does not fully exploit the features of RBMs as architectures that naturally describe quantum states in a nonlocal manner~\cite{melko2019restricted,RevModPhys.93.045003}. Thus, we choose to build a paradigm of direct analysis of the spatial complexity of long-range RBMs.

In this paper, we analyze the efficiency of long-range RBM state representation for 1D quantum spin systems. Our procedure is as follows:
\begin{enumerate}
    \item In Sec.~\ref{sec:LRFD}, we generalize the RBM wave-function ansatz to an infinitely-many-hidden-node regime and define a subset of generic RBM states---the long-range-fast-decay (LRFD) RBM states, whose parameter conditions constrain the nonlocal interactions between spins (visible nodes) and virtual particles (hidden nodes).
    \item In Sec.~\ref{sec:trunc}, we derive an upper bound on truncation errors associated with two measures of state differences for the sequence of truncated LRFD RBM states. One measure is the $l_{2}$-norm of the state-vector difference and the other is a Hermitian-operator-based expectation-value difference. 
    \item In Sec.~\ref{sec:scaling}, we identify the dependence of the spatial complexity for LRFD RBMs in state approximation on the decaying rates specified in the nonlocal interaction pattern. 
    \item In Sec.~\ref{sec:ground}, we provide numerical evidence supporting a conjecture that the ground states of a wide range of 1D quantum spin systems, including some critical systems with logarithmic entanglement entropy, can be approximated by LRFD RBMs with the scaling of the spatial complexity being at most polynomial in both the system size and the inverse of approximation errors. 
    \item In Sec.~\ref{sec:ground}, we also provide the relations between multiple typical state manifolds through which the importance of the concept of LRFD RBMs in efficiency analysis for state representation theory is manifested.
\end{enumerate}

Our results offer evidence for the utility of RBMs in cases where other state parameterizations, such as matrix product states (MPSs), become less efficient. Our work actually proposes a paradigm of doing complexity analysis for general long-range RBMs, rather than limited to short-range or sparse RBMs, and naturally yields a further classification of this manifold based on the complexity scaling.

We find that the nonlocal structure of LRFD RBMs can be characterized by two conditions. These conditions are each determined by bounds
associated with two degrees of freedom, defined within a framework of \textit{levels} that is depicted in Fig.~\subref*{fig:network}. One of the two degrees of freedom is a single-level decaying factor resembling localized orbitals and encoding information about correlations between spins (Sec.~\ref{sec:correlation}). The second is a level-decay factor, which has a significant influence on the complexity of the RBMs (Sec.~\ref{sec:scaling}). 

This paradigm and our characterization of the nonlocal structures may promote the understanding of the natural measure of complexities for quantum many-body states described by RBMs and may be generalizable to higher-dimensional systems and to deep neural-network quantum states.


\section{The restricted Boltzmann machine as a wave-function ansatz} \label{sec:wfantz}

\captionsetup[subfigure]{position=top,singlelinecheck=off,justification=raggedright}
\begin{figure}[tbp]
\centering
 \subfloat[]{\includegraphics[width=0.225\textwidth]{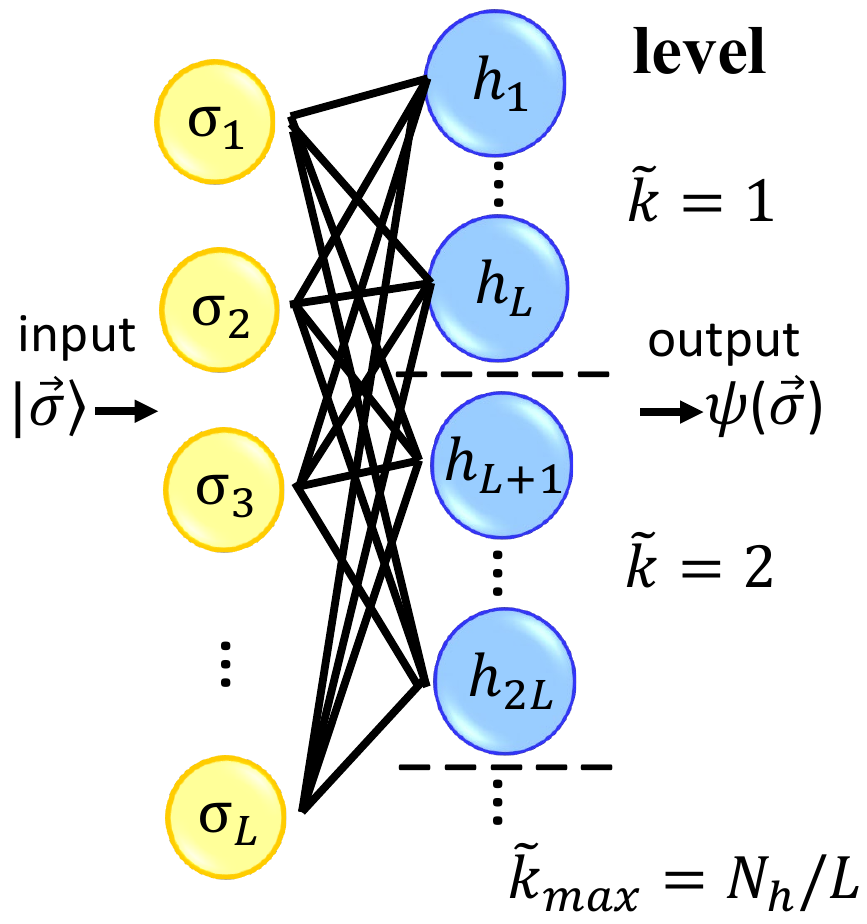} \label{fig:network}}
 \subfloat[]{\includegraphics[width=0.255\textwidth,height=0.23\textwidth]{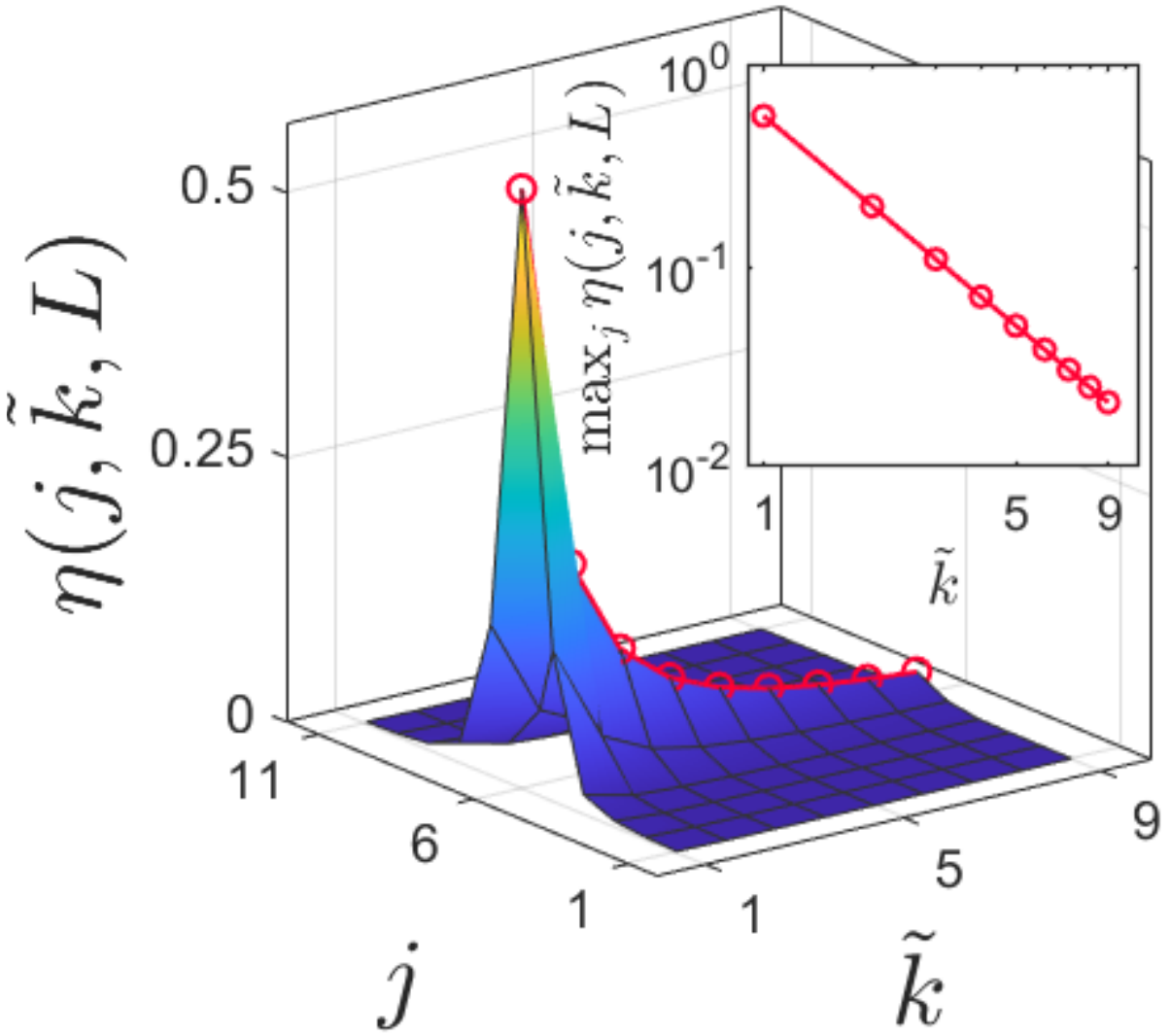} \label{fig:3Dstandard}}
  \caption{(a) Network structure of RBMs as a wave-function ansatz for 1D quantum spin systems. Long-range RBMs usually implies full connectivity between the visible and hidden layer. The hidden layer is divided into $N_{h}/L$ levels, each containing $L$ hidden nodes. (b) Importance measure $\eta(j,\tilde{k},L)$ for a LRFD RBM with translational symmetry. The RBM is constructed as Eqs.~(\ref{eq:LRFDW})--(\ref{eq:LRFDa}) show, where $\lambda(\tilde{k})=\tilde{k}^{-\alpha_{P}}$, $\mu(r)=\frac{1}{2}\delta_{Q}r^{-\alpha_{Q}}$ for $r\neq 0$, $\mu(0)=\delta_{Q}=0.5$, $\alpha_{P}=0.75$, $\alpha_{Q}=1.5$, $c_{w}=1+i$, $c_{b}=0$, $a_{0}=0$ and $L=11$. The inset plots the decay of the maximum of $\eta(j,\tilde{k},L)$ among all $j$ at each level with increasing $\tilde{k}$ on a log-log scale. The linearity of the curve reveals a power-law decaying of the ``ridge'' (red circles) of the 3D structure.}
\end{figure}

We use the RBM as a wave-function ansatz for 1D quantum many-body spin-$\frac{1}{2}$ systems~\cite{sarma2019machine,carleo2017solving,melko2019restricted}. The RBM usually works as the building block for understanding and training deeper networks because of its relatively simple structure for inference and its power in parametric modeling as a universal approximator for discrete distribution~\cite{6796877}. As basic constructs of deep NNs, the RBMs have two \textit{layers}. The first layer (a \textit{visible} layer) represents a spin configuration $\vec{\sigma}$ in the usual way. Here, the vector $\vec{\sigma}=(\sigma_{1},\ldots,\sigma_{L})$ represents a system of $L$ spins with $\sigma_{j}=\pm 1$ for $j=1,\ldots,L$. The second layer is a \textit{hidden} layer. It is composed of $N_{h}$ nodes, denoted by a vector $\vec{h}=(h_{1},\ldots,h_{N_{h}})$ with $h_{k}=\pm 1$ for $k=1,\ldots,N_{h}$. The $h_{k}$'s are introduced as auxiliary particles in the probability model; they play roles similar to those of virtual particles in the valence-bond picture for MPSs~\cite{RevModPhys.93.045003,RevModPhys.82.277}.

Given a specific spin configuration $\vec{\sigma}$, the RBM outputs the corresponding wave-function amplitude
\begin{IEEEeqnarray}{rCl}
\psi(\vec{\sigma})
&=& 2^{-N_{h}} \sum_{\{ \vec{h}: h_{k}=\pm 1 \}} \exp \Big( \sum_{j=1}^{L}a_{j}\sigma_{j}+\sum_{k=1}^{N_{h}}b_{k}h_{k} \nonumber\\
&& +\sum_{1\le j\le L,1\le k\le N_{h}, j,k \in \mathbb{N}}W_{j,k}\sigma_{j}h_{k} \Big)  \\
&=& \prod_{j=1}^{L}e^{a_{j}\sigma_{j}}\prod_{k=1}^{N_{h}}\cosh(b_{k}+\sum_{j=1}^{L}\sigma_{j}W_{j,k}).
\end{IEEEeqnarray} 
Here, $a_{j}$ and $b_{k}$ are the bias parameters for the $j$-th spin and $k$-th hidden node, respectively, $W_{j,k}$ is a weight parameter describing the interlayer interaction between the $j$-th spin and $k$-th hidden node, and $\mathbb{N}$ denotes the set of all natural numbers. The $a_{j}$, $b_{k}$ and $W_{j,k}$ are complex numbers. All such amplitudes defined on the computational basis yield a quantum state vector $|\Psi\rangle = \sum_{\vec{\sigma}}\psi(\vec{\sigma})|\vec{\sigma}\rangle$, where the summation is over all $2^{L}$ spin configurations. It is remarkable that we adopt the RBM form with a factor of  $2^{-N_{h}}$. This choice allows us to use infinitely many hidden nodes $h_{k}$ as long as $b_{k}$ and $W_{j,k}$ decay sufficiently fast to ensure the convergence of $\psi(\vec{\sigma})$ as $N_{h}\to \infty$ for fixed system sizes $L$. In other words, it ensures that adding hidden nodes with associated parameters ($b_{k}$ and $W_{j,k}$) being zero will not change the value of the wave function. This choice will facilitate the asymptotic analysis as shown below.

As mentioned in Sec.~\ref{sec:QML3intro}, the RBMs solved by relevant ML algorithms to approximate target states often feature a long-range form and a fast parameter decay. As more hidden nodes are added to the network, the RBM can capture higher-order correlations between spins~\cite{carleo2017solving}, thus leading to higher accuracy in approximation. The parameter decay is manifested by the decay of weight parameters $W_{j,k}$ with an increasing index separation $|j-k|$ as well as the decay of $b_{k}$ with increasing $k$. 

In this work, we assume $N_{h}$ to be an integer multiple of $L$ which will facilitate the scaling analysis, especially for translationally invariant systems. When $N_{h}$ is not an integer multiple of $L$, we can simply fill the last fragment with hidden nodes associated with zero-value parameters without influencing the wave-function values. We divide the hidden layer into multiple \textit{levels}, each of which contains $L$ hidden nodes (Fig.~\subref*{fig:network}). Thus, there are totally $N_{h}/L$ levels while the ratio $N_{h}/L$ is called the hidden-unit density in some references~\cite{carleo2017solving}. We will show that hidden nodes at the same level can capture the correlation of the same order between spins by performing an algorithm to reorder all hidden nodes for general RBMs. This point will be further clarified when we use the RBM form with translational symmetry to represent the ground states of 1D translationally invariant quantum systems as shown below~\cite{carleo2017solving}.

One example of the quantum states that can be exactly represented by short-range RBMs~\cite{deng2017machine,deng2017quantum} is the 1D symmetry-protected topological (SPT) cluster state. The Hamiltonian of the SPT cluster system is defined on a 1D $L$-site lattice with periodic boundary conditions as $\hat{H}_{\textrm{cluster}} = -\sum_{j=1}^{L}\hat{\sigma}_{j-1}^{z}\hat{\sigma}_{j}^{x}\hat{\sigma}_{j+1}^{z}$, where $\hat{\sigma}^{x}$ and $\hat{\sigma}^{z}$ are Pauli matrices. A conventional $r_{0}$-range RBM is defined as an RBM satisfying $W_{j,k}=0$ for any $|j-k|>r_{0}$. A short-range RBM usually refers to an $r_{0}$-range RBM with $r_{0}$ being a small constant independent of the system size $L$. It was shown in Ref.~\cite{deng2017machine,deng2017quantum} that the ground state of $\hat{H}_{\textrm{cluster}}$ can be exactly represented by a $1$-range RBM with $L$ hidden nodes defined as: 
\begin{IEEEeqnarray}{rCl} \label{eq:SPT}
&& a_{j} = 0 \textrm{ (for any } j\in\{1,2,\ldots,L\}), \nonumber\\
&& b_{k} = i\pi/4, W_{k-1,k}=i\pi/2, W_{k,k} = 3i\pi/4, \nonumber\\
&& W_{k+1,k} = i\pi/4 \textrm{ (for any } k\in\{1,2,\ldots,L\}), \nonumber\\
&& W_{j,k} = 0 \textrm{ (for } |j-k|>1),
\end{IEEEeqnarray}
by using the stabilizer nature of the system to decrease the number of equation constraints for parameters from exponential to linear in $L$. Using our language of levels, this RBM just has one level and its weight parameters at this single level have a support of very short length which is a manifestation of its quantum entanglement satisfying an area law. Moreover, the translational symmetry of the system is inherited by the RBM form. The parameter patterns of this RBM also have a translational symmetry, which means that its parameters for different hidden nodes can be generated by the action of a translational-symmetry transformation operator on those for a single hidden node~\cite{carleo2017solving}.      

Inspired by the extensibility of the system of equations~(\ref{eq:SPT}) with growing system sizes and considering the need to capture higher-order correlations between spins~\cite{carleo2017solving} and stronger quantum entanglement between subsystem blocks~\cite{deng2017quantum}, we expect that the RBM representation of general quantum states has multiple, possibly infinitely many, levels and the length of the support of weight parameters at each level may increase from a small constant to the maximum length $L$. This motivates us to analyze generic long-range RBMs with properly specified nonlocal interactions between spins and hidden nodes (virtual particles).

\subsection{Long-range-fast-decay RBMs} \label{sec:LRFD}

We now discuss aspects of the nonlocal structure of LRFD RBMs that were summarized at the end of Sec.~\ref{sec:QML3intro}. This leads to specific definitions of the two conditions that were mentioned there.

We begin by generalizing the RBM wave-function ansatz to an infinitely-many-hidden-node regime. An RBM state $|\Psi^{(L,\infty)}\rangle$ with infinitely many hidden nodes and a system size $L$ can be defined as 

\begin{IEEEeqnarray}{rCl}
|\Psi^{(L,\infty)}\rangle = \sum_{\vec{\sigma}}\psi^{(L,\infty)}(\vec{\sigma})|\vec{\sigma}\rangle,
\end{IEEEeqnarray}  
where 
\begin{IEEEeqnarray}{rCl} \label{eq:RBMinfty}
\psi^{(L,\infty)}(\vec{\sigma})
= \prod_{j=1}^{L}e^{a_{j}^{(L)}\sigma_{j}}\prod_{k=1}^{\infty}\cosh(b_{k}^{(L)}+\sum_{j=1}^{L}\sigma_{j}W_{j,k}^{(L)}).  \nonumber\\
\end{IEEEeqnarray}  
Its corresponding truncated-RBM sequence is defined as 
$\{ |\Psi^{(L,N_{h})}\rangle \}$, where
\begin{IEEEeqnarray}{rCl}
|\Psi^{(L,N_{h})}\rangle = \sum_{\vec{\sigma}}\psi^{(L,N_{h})}(\vec{\sigma})|\vec{\sigma}\rangle
\end{IEEEeqnarray} 
and
\begin{IEEEeqnarray}{rCl}
\psi^{(L,N_{h})}(\vec{\sigma})
= \prod_{j=1}^{L}e^{a_{j}^{(L)}\sigma_{j}}\prod_{k=1}^{N_{h}}\cosh(b_{k}^{(L)}+\sum_{j=1}^{L}\sigma_{j}W_{j,k}^{(L)}) \nonumber\\
\end{IEEEeqnarray} 
is constructed by removing the hyperbolic cosine terms with $k\ge N_{h}+1$ from $\psi^{(L,\infty)}(\vec{\sigma})$.


Then, we define a subset of generic RBM states with infinitely many hidden nodes---long-range-fast-decay (LRFD) RBM states---as the RBMs whose parameters satisfy the following two conditions.

\begin{condition}[\textit{boundedness of $W_{j,k}$}] \label{cond:W}
There exists an $L$-independent integer $\tilde{k}_{s}\in \mathbb{N}$ and three nonnegative monotonically decreasing real functions $\lambda_{R}(\tilde{k})$, $\lambda_{I}(\tilde{k})$ and $\mu(r)$ such that, after a reordering of all hidden nodes, for all $k>\tilde{k}_{s}L$,
\begin{IEEEeqnarray}{rCl}
| \operatorname{Re}(W_{j,k}^{(L)}) | 
&\le& \lambda_{R}(\tilde{k}) \mu(|j-j_{\textrm{c}}|_{\textrm{circ}}),\\
| \operatorname{Im}(W_{j,k}^{(L)}) | 
&\le& \lambda_{I}(\tilde{k}) \mu(|j-j_{\textrm{c}}|_{\textrm{circ}}),
\end{IEEEeqnarray}
where $\tilde{k}\in \{ 1,2,\ldots,N_{h}/L \}$ designates the numerical index of levels; $j_{\textrm{c}}$, the center spin for the $k$-th hidden node, denotes the site index of the spin with which the interaction of the $k$-th hidden node reaches its maximum among all $j\in\{ 1,2,\ldots,L\}$; $|m|_{\textrm{circ}}=\min\{m, L-m\}$ in accordance with the periodic boundary conditions; and $r\in\{ 0,1,\ldots,(L-1)/2 \}$ denotes the distance between $j$ and $j_{c}$ assuming $L$ is odd without influencing the validity of the following asymptotic analysis. The functions $\lambda_{R}(\tilde{k})$, $\lambda_{I}(\tilde{k})$ and $\mu(r)$ satisfy the conditions that there exist finite $L$-independent nonnegative constants $P_{0}$ and $\mu_{0}$ such that
\begin{IEEEeqnarray}{rCl}
& \sum_{\tilde{k}=\tilde{k}_{s}+1}^{\infty}
\Big( \lambda_{R}^{2}(\tilde{k}) + \beta_{1}^{2}\lambda_{I}^{2}(\tilde{k}) \Big) = P_{0}<\infty, & \label{eq:P0}\\
& \mu(r) \le \mu_{0}<\infty \quad (\textrm{for all } r \ge 0),& \label{eq:mu0}
\end{IEEEeqnarray}
where $\beta_{1}=3\sqrt{2\ln2}/\pi$ is found in the convergence proof given in Appendix~\ref{sec:proofconverge}.
\end{condition}

We provide an interpretation of each new variable as follows. After a reordering of all hidden nodes which is usually associated with sorting based on the value of $| \operatorname{Re}(W_{j,k}^{(L)}) |^{2} + \beta_{1}^{2} | \operatorname{Im}(W_{j,k}^{(L)}) |^{2}$, starting from the level $\tilde{k}_{s}+1$, $\tilde{k}=\tilde{k}(k)$ and $j_{\textrm{c}}=j_{\textrm{c}}(k)$ are both functions of $k$ and the correspondence between the pair $(\tilde{k},j_{\textrm{c}})$ and $k$ is a bijective map. It means that every hidden node with $k>\tilde{k}_{s}L$ is associated with a unique pair and thus can be uniquely positioned in the RBM network after the reordering (Fig.~\subref*{fig:network}). The hidden nodes capturing the correlation of the same order between spins are grouped at the same level so that the new indices of these hidden nodes characterized by the pair $(\tilde{k},j_{\textrm{c}})$ actually manifest the level of correlations. This characterization can also facilitate a symmetry manifestation for quantum states holding translational symmetry. The reordering step is to solve the problem that ML algorithms with a stochastic nature are often unable to automatically group the hidden nodes according to level stratification and their site positions usually exhibit randomness. The condition $\tilde{k}_{s}=0$ implies that all hidden nodes satisfy the boundedness conditions so that the level stratification can be applied to the whole hidden layer (Fig.~\subref*{fig:network}).

\begin{condition}[\textit{boundedness of $b_{k}$}]
\label{cond:b}
After the same reordering of all hidden nodes that is described in Condition~\ref{cond:W}, for all $k>\tilde{k}_{s}L$,
\begin{IEEEeqnarray}{rCl}
| \operatorname{Re}(b_{k}^{(L)}) | 
&\le& \lambda_{R}(\tilde{k})\mu(0),\\
| \operatorname{Im}(b_{k}^{(L)}) | 
&\le& \lambda_{I}(\tilde{k})\mu(0).
\end{IEEEeqnarray}
\end{condition}

The definition of LRFD RBMs should be understood from the point of view of state manifolds~\cite{huang2017complexity,RevModPhys.93.045003}. A state manifold for quantum many-body states usually refers to a subspace of the whole Hilbert space spanned by a parameterized wave-function family~\cite{huang2017complexity}, thus is a set containing specific types of quantum states. So the manifold of LRFD RBMs can be defined as a space spanned by all parameterized wave functions, every one of which belongs to a quantum-state sequence associated with a varying system size and satisfying the above Condition~\ref{cond:W} and ~\ref{cond:b}. One LRFD-RBM state refers to an element in this manifold. So this definition is in the same spirit as the definition of MPSs with different scaling laws~\cite{RevModPhys.93.045003,RevModPhys.82.277}.

Condition~\ref{cond:W} gives an upper bound on the magnitude of RBM weight parameters and actually provides a description of the nonlocal interaction between spins and hidden nodes (virtual particles). It requires that $| {\operatorname{Re}(W_{j,k}^{(L)})} |$ and $| {\operatorname{Im}(W_{j,k}^{(L)})} |$ are upper bounded, respectively, by the products $\lambda_{R}(\tilde{k})\mu(r)$ and $\lambda_{I}(\tilde{k})\mu(r)$. The monotonically decreasing functions $\lambda_{R}(\tilde{k})$ and $\lambda_{I}(\tilde{k})$ can be regarded as level-decay factors, while $\mu(r)$ is a factor describing the decay due to the increase of the distance between the spin-site index ($j$)  and the corresponding spin-site index of the center spin ($j_{\textrm{c}}(k)$) for the $k$-th hidden node. The function $\mu(r)$ has a localization feature and resembles a single-modal localized orbital in the physics of periodic potentials, such as Wannier modes~\cite{PhysRevX.8.031087}, which can be reflected by its monotonically decreasing with increasing $r$. So this description can effectively capture the parameter decays induced by both the level increase and the growth of system size, providing two degrees of freedom in characterizing the nonlocal interaction pattern. The separate treatments for the real and imaginary parts originate from their inequivalent positions in the RBM wave-function form, which is shown in Appendix~\ref{sec:proofconverge}.

Condition~\ref{cond:b} implies that the contribution of $b_{k}$-related terms can be upper bounded by the largest $W_{j,k}$-related terms at each level so that the $W_{j,k}$ weight parameters play a dominant role in the asymptotic analysis (Appendix~\ref{sec:proofconverge}). Since there is often a degree of freedom in choosing the value of $\mu(0)$, Condition~\ref{cond:b} can be satisfied for a wide range of RBM states.

Conditions~\ref{cond:W} and \ref{cond:b} are proposed to ensure the convergence of the state vector (Eq.~(\ref{eq:RBMinfty})) and provide a clear quantification for the rate of parameter decays, on the basis of which a complexity analysis can be conducted. A rigorous proof of the convergence of the state vector when Conditions~\ref{cond:W} and \ref{cond:b} are satisfied is given in Appendix~\ref{sec:proofconverge}. This proof is important not only because it ensures that the generalization of RBMs to an infinitely-many-hidden-node regime makes sense by defining them as the limits of some infinite sequences, but also because it introduces the key mathematical tricks and concepts that are necessary for analyzing the effects of truncations.

\captionsetup[subfigure]{position=top,singlelinecheck=off,justification=raggedright}
\begin{figure}[tbp]
\centering
 \subfloat[]{\includegraphics[width=0.245\textwidth]{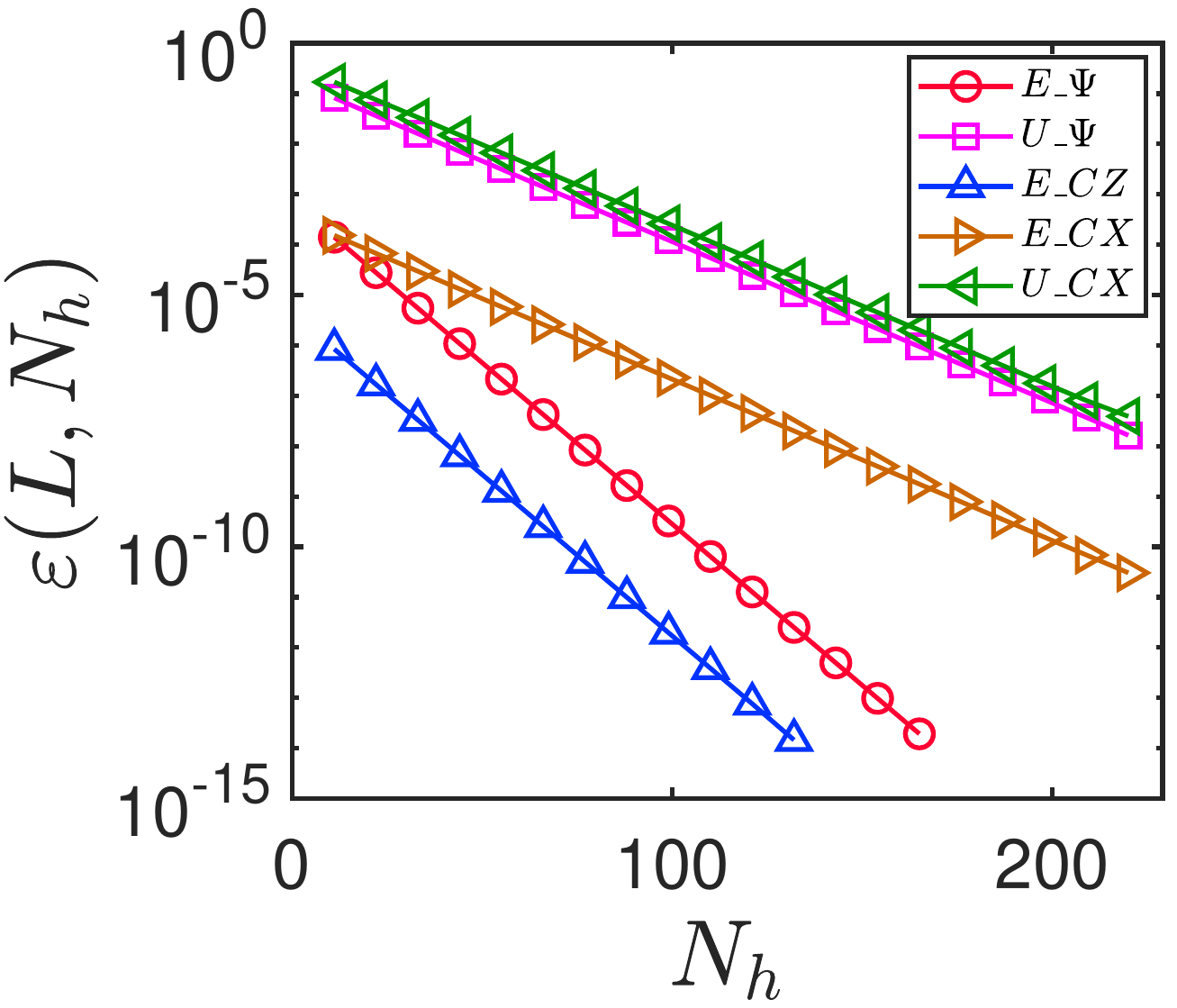} \label{fig:SPTexp}}
 \subfloat[]{\includegraphics[width=0.245\textwidth]{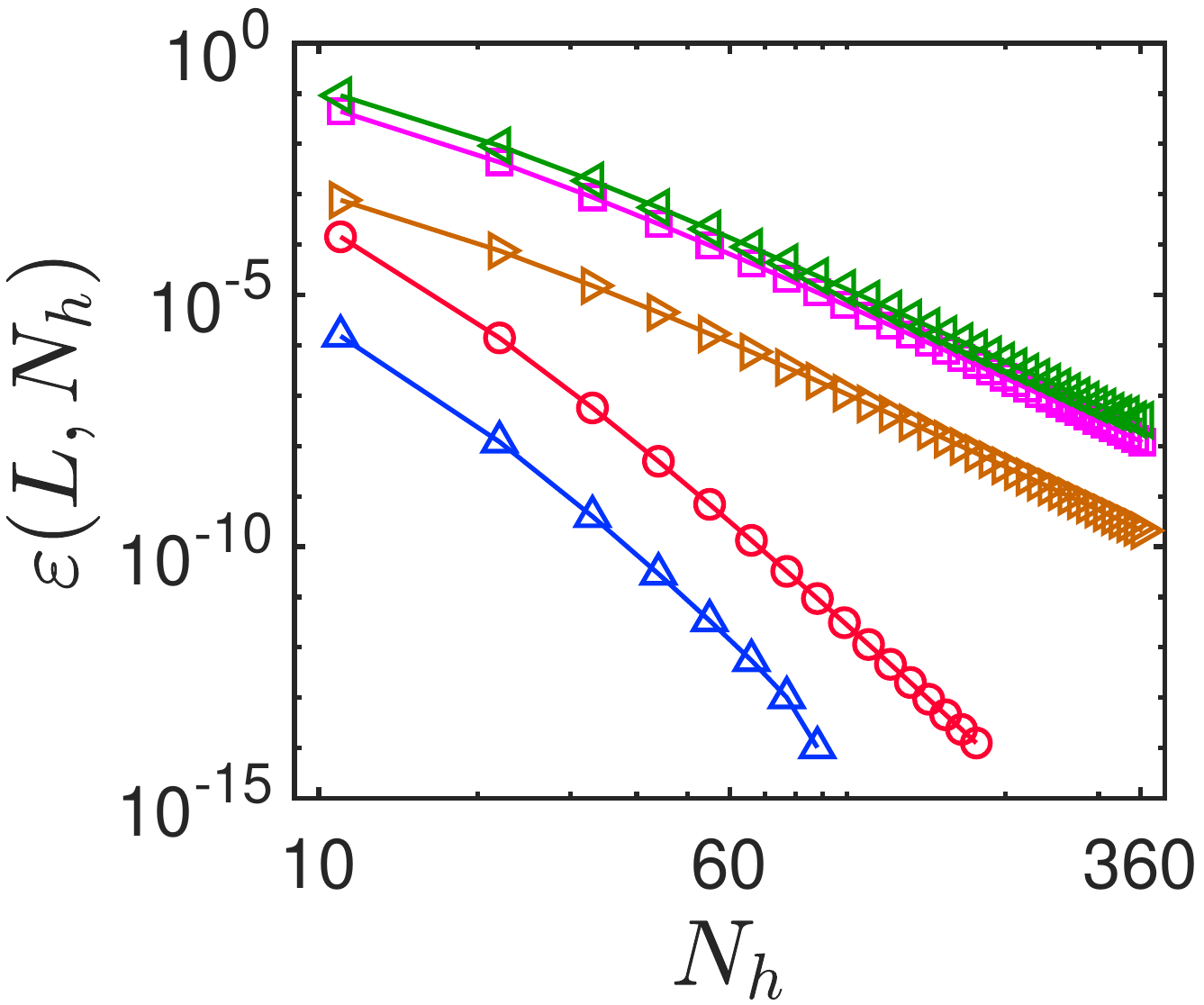} \label{fig:SPTpower}}\\
\subfloat[]{\includegraphics[width=0.23\textwidth]{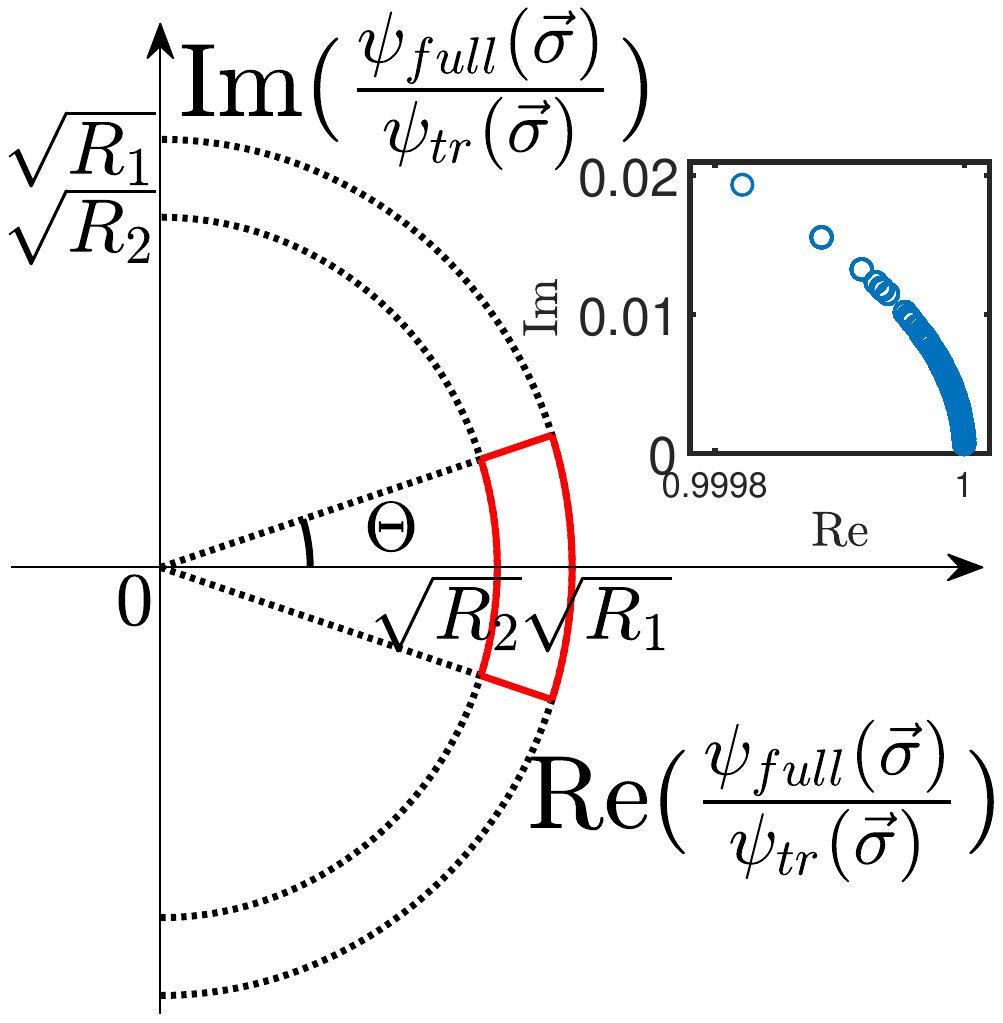} \label{fig:ratio}}
\subfloat[]{\includegraphics[width=0.258\textwidth]{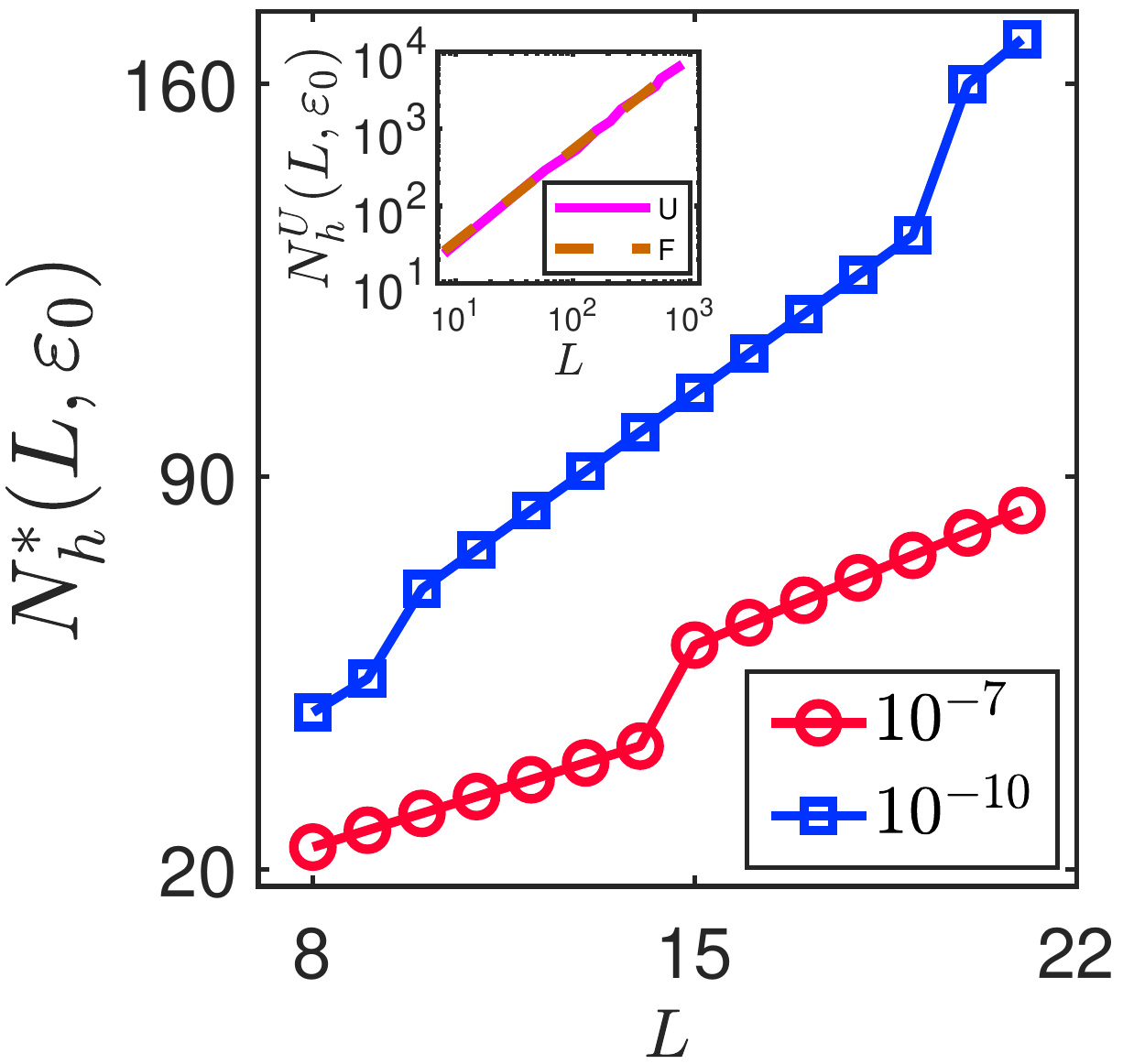} \label{fig:NhvsL}}
\caption{(a)(b) Comparison between the exact and estimated truncation errors $\varepsilon(L,N_{h})$ as a function of $N_{h}$ with fixed $L$ for 1D SPT cluster states with a perturbation part. $E\_\Psi$: exact values, first-type truncation errors; $U\_\Psi$: upper-bound-based estimation, first-type; $E\_CZ$: exact, second-type, $\hat{B}=\hat{\sigma}_{1}^{z}\hat{\sigma}_{2}^{z}$; $E\_CX$: exact, second-type, $\hat{B}=\hat{\sigma}_{1}^{x}\hat{\sigma}_{2}^{x}$; $U\_CX$: upper-bound-based estimation, second-type, $\hat{B}=\hat{\sigma}_{1}^{x}\hat{\sigma}_{2}^{x}$.   The perturbation part is constructed as Eqs.~(\ref{eq:LRFDW})--(\ref{eq:LRFDa}) show. $\mu(r)=\frac{1}{2}\delta_{Q}r^{-\alpha_{Q}}$ for $r\neq 0$, $\mu(0)=\delta_{Q}=0.1$, $\alpha_{Q}=3$, $c_{w}=c_{b}=1+i$, $a_{0}=0$ and $L=11$. (a) Exponential decay of $\lambda(\tilde{k})$ (vertical axis: log scale). $\lambda(\tilde{k})=0.2\delta_{P}^{-(\tilde{k}-1)}$ with $\delta_{P}=1.5$. (b) Power-law decay of $\lambda(\tilde{k})$ (on a log-log scale). $\lambda(\tilde{k})=\tilde{k}^{-\alpha_{P}}$ with $\alpha_{P}=3$.  (c) Schematic interpretation of variables used in proofs. The inset in (c) shows the distribution of data points of the ratio $\psi^{(L,\infty)}(\vec{\sigma}) / \psi^{(L,N_{h})}(\vec{\sigma})$ with $N_{h}=L$, which corresponds to a truncation removing hidden nodes starting from the second level. The parameter setting is the same as that in (b). The data points are all localized in the neighborhood of $z=1$ in the complex plane enclosed by the red solid curves in (c) as we analyzed.  (d) Scaling of $N_{h}^{\ast}(L,\varepsilon_{0})$ in $L$ for two fixed values of the first-type truncation errors $\varepsilon_{0}$ with the same parameter setting as in (b). Red circle: $\varepsilon_{0}=10^{-7}$; Blue square: $\varepsilon_{0}=10^{-10}$. The inset in (d) shows the scaling of $N_{h}$ estimated based on our upper bounds with $\varepsilon_{0}=10^{-3}$. Magenta solid: using exact values of upper bounds; Brown dashed: using leading-order estimations. The two curves almost coincide.}
\end{figure}

The core idea of the proof is that we can prove the sequence $\{ \psi^{(L,nL)}(\vec{\sigma}): n\in \mathbb{N} \}$ is a Cauchy sequence in the field of complex numbers $\mathbb{C}$~\cite{horn2012matrixcauchyseq}. This proof is inspired by the fact that, when $b_{k}^{(L)}$ and $W_{j,k}^{(L)}$ decay sufficiently fast, the complex-valued ratio $\psi^{(L,(n+m)L)}(\vec{\sigma}) / \psi^{(L,nL)}(\vec{\sigma})$ will quickly fall into the neighborhood of the point $z=1$ in the complex plane as $n$ increases. So we derive an upper and lower bound on the ratio's modulus $|\psi^{(L,(n+m)L)}(\vec{\sigma}) / \psi^{(L,nL)}(\vec{\sigma})|$ which converge to $1$ and an upper bound on the magnitude of its argument $| \arg \big( \psi^{(L,(n+m)L)}(\vec{\sigma}) / \psi^{(L,nL)}(\vec{\sigma}) \big) |$ which converges to $0$ as $n$ increases.  Then we show that the corresponding magnitude sequence $\{ | \psi^{(L,nL)}(\vec{\sigma}) | \}$ and the argument sequence $\{ \arg( \psi^{(L,nL)}(\vec{\sigma}) ) \}$ are Cauchy sequences in the field of real numbers $\mathbb{R}$, thus $\{ \psi^{(L,nL)}(\vec{\sigma})  \}$ is a Cauchy sequence in $\mathbb{C}$.

\subsection{Effects of wave-function truncation for fixed system sizes} \label{sec:trunc}

We derive upper bounds on truncation errors associated with two measures of state differences for the sequence of truncated LRFD RBM states. Define $\varepsilon(L,N_{h})$ to be a specific type of truncation error for using $|\Psi^{(L,N_{h})}\rangle$ to approximate $|\Psi^{(L,\infty)}\rangle$. 

A natural measure of state differences is the square of the $l_{2}$-norm~\cite{horn2012matrixvnorm} of the state-vector difference, $\Vert |\tilde{\Psi}^{(L,\infty)} \rangle - |\tilde{\Psi}^{(L,N_{h})} \rangle \Vert_{2}^{2} $, where the tilde symbol is used to represent corresponding states after a normalization operation. It is remarkable that the RBM wave-function ansatz is not automatically normalized and an estimation of the normalization factor $\langle \Psi | \Psi \rangle$ is important and often tricky as shown in Appendix~\ref{sec:proofbound}. This measure of truncation errors is adopted in fundamental works about the faithfulness and efficiency of other wave-function ans{\" a}tze, such as MPSs~\cite{PhysRevB.73.094423,10.5555/2011832.2011833}. So it allows us to make a direct comparison between the efficiencies of RBMs and other state representations. 

A second measure of state differences is a Hermitian-operator-based expectation-value difference defined as
\begin{IEEEeqnarray}{rCl}
&& | \langle\hat{B}\rangle^{(L,\infty)} - \langle\hat{B}\rangle^{(L,N_{h})} | \nonumber\\
&=& | \langle \tilde{\Psi}^{(L,\infty)} | \hat{B} |\tilde{\Psi}^{(L,\infty)} \rangle - \langle \tilde{\Psi}^{(L,N_{h})} | \hat{B} |\tilde{\Psi}^{(L,N_{h})} \rangle | . 
\end{IEEEeqnarray}
Here, $\hat{B}$ can be any Hermitian operator of the form $\hat{B} = \bigotimes_{j=1}^{L} \hat{\sigma}_{j}^{(m_{j})}$, where $\bigotimes$ is the tensor product symbol, $m_{j}\in\{0,1,2,3\}$, $\hat{\sigma}_{j}^{(0)}=I_{2\times 2}$ is a $2$-by-$2$ identity matrix, and $\{ \hat{\sigma}_{j}^{(1)}, \hat{\sigma}_{j}^{(2)}, \hat{\sigma}_{j}^{(3)} \}$ denote the Pauli matrices. We also use this measure as $\{ \hat{\sigma}_{j}^{(m)}: m=0,1,2,3\}$ is a complete basis set for the local Hilbert space for the $j$-th spin and a wide range of typical physical observables, such as spin correlations and total energy, correspond to Hermitian operators of such type or linear combinations of polynomially many such operators.

Then we can prove a lemma which provides upper bounds on truncation errors of the above two types for the sequence of truncated LRFD RBM states.

\begin{lemma}[\textit{upper bounds on truncation errors}] 
\label{lemma:bound}
For LRFD RBMs satisfying Conditions~\ref{cond:W} and \ref{cond:b}, after the same reordering of all hidden nodes described in Condition~\ref{cond:W}, there exists $n_{\Theta}(L) > \tilde{k}_{s}$ such that, for all $N_{h}>n_{\Theta}(L)L$,
\begin{IEEEeqnarray}{rCl}
\Vert |\tilde{\Psi}^{(L,\infty)} \rangle - |\tilde{\Psi}^{(L,N_{h})} \rangle \Vert_{2}^{2} &\le& F_{1}\Big( LQ(L) P(N_{h}/L) \Big), \label{eq:errorPsi}\\
| \langle\hat{B}\rangle^{(L,\infty)} - \langle\hat{B}\rangle^{(L,N_{h})} |
&\le& F_{2}\Big( LQ(L)P(N_{h}/L) \Big), \label{eq:errorB}
\end{IEEEeqnarray}
where
\begin{IEEEeqnarray}{rCl}
F_{1}(x) 
&=& 2-2 \exp [ -2(1+\beta_{1}^{2})x ] \cos(4\beta_{2}x) \label{eq:F1x} \\
&=& c_{1}x+\mathcal{O}(x^{2}) \quad (\textrm{as } x\to 0), \\
F_{2}(x) 
&=& \max\{ |\exp(4x)-1|, |1-\exp(-4\beta_{1}^{2}x)| \} \nonumber\\
&& + \max\{ \Big[ \exp(8x)-2\exp(4x)\cos(8\beta_{2}x) + 1 \Big]^{1/2}, \nonumber\\
&& \Big[ \exp(-8\beta_{1}^{2}x)-2\exp(-4\beta_{1}^{2}x)\cos(8\beta_{2}x) + 1 \Big]^{1/2} \} \label{eq:F2x} \nonumber\\
\\
&=& c_{2}x+\mathcal{O}(x^{3/2})  \quad (\textrm{as } x\to 0).   \\
P(m) &=& \sum_{\tilde{k}=m+1}^{\infty} \lambda^{2}(\tilde{k}) 
 \quad (m\ge \tilde{k}_{s}, m\in \mathbb{N}), \label{eq:Px} \\
Q(L) &=&  \Big( \sum_{r=0}^{(L-1)/2}\mu(r) \Big)^{2}, \label{eq:Qx}
\end{IEEEeqnarray}
the relevant constants are $\beta_{2} = 3\sqrt{3}/\pi$, $c_{1} = 4(1+\beta_{1}^{2})$ and $c_{2} = 4\beta_{1}^{2}+4(\beta_{1}^{4}+4\beta_{2}^{2})^{1/2}$, $n_{\Theta}(L)$ can be estimated by inequality (\ref{eq:nTheta}), and we have assumed that $\lambda_{R}(\tilde{k}) = \lambda_{I}(\tilde{k}) = \lambda(\tilde{k})$ for simplicity which holds throughout the following discussion. 
\end{lemma}

The proof is given in Appendix~\ref{sec:proofbound} which uses arguments similar to those described in the proof for the convergence of LRFD RBMs. Based on the intuition that the ratio $\psi^{(L,\infty)}(\vec{\sigma}) / \psi^{(L,N_{h})}(\vec{\sigma})$ will fastly converge to $1$ with increasing $N_{h}$, we derive an upper bound $\sqrt{R_{1}}$ and a lower bound $\sqrt{R_{2}}$ on the ratio's modulus $| \psi^{(L,\infty)}(\vec{\sigma}) / \psi^{(L,N_{h})}(\vec{\sigma}) |$ and an upper bound $\Theta$ on the magnitude of its argument $| \arg \big( \psi^{(L,\infty)}(\vec{\sigma}) / \psi^{(L,N_{h})}(\vec{\sigma}) \big) |$. The two types of truncation errors can be upper bounded using these three variables and the two upper bounds can be finally expressed as functions ($F_{1}(x)$ and $F_{2}(x)$) of $LQ(L) P(N_{h}/L)$ which decreases to zero with increasing $N_{h}$ and fixed $L$. The idea of the proof is shown schematically in Fig.~\subref*{fig:ratio}. 

Based on our description of the nonlocal interactions between spins and virtual particles and using the language of levels, $P(N_{h}/L)$ is a summation of all level-decay factors for hidden nodes at levels starting from $\tilde{k}=N_{h}/L+1$ to $\tilde{k}=\infty$, while $Q(L)$ corresponds to the localized ``orbitals'' at every single level and contributes a factor reflecting the pure influence of system-size growing regardless of levels. The two different types of truncation errors correspond to two different forms of the function $F(x)$, but both of them are analytic at the point $x=0$. 

We give the scaling of truncation errors in $N_{h}$ as below. It can be obtained that, if $Q(x)=\mathcal{O}(q(x))$ as $x\to \infty$, $P(x)=\mathcal{O}(1/p_{d}(x))$ as $x\to \infty$, and $F(x)=\mathcal{O}(f(x))$ as $x\to 0$, then 
\begin{IEEEeqnarray}{rCl} \label{eq:epsvsNh}
\varepsilon(L,N_{h})=\mathcal{O}(f(\frac{L\, q(L)}{p_{d}(N_{h}/L)})) \quad (\textrm{as } N_{h}\to \infty).
\end{IEEEeqnarray}

Our construction of LRFD RBMs and theoretical analysis of the truncation errors can be further clarified with results from numerical computations. We can construct LRFD RBMs with translation symmetry whose parameters exactly satisfy
\begin{IEEEeqnarray}{rCl}
W_{j,k}^{(L)}
&=& c_{w}\lambda(\tilde{k}) \mu(|j-j_{\textrm{c}}|_{\textrm{circ}}), \label{eq:LRFDW}\\
b_{k}^{(L)} &=& c_{b}\lambda(\tilde{k})\mu(0), \label{eq:LRFDb}\\
a_{j}^{(L)} &=& a_{0} \label{eq:LRFDa}
\end{IEEEeqnarray}
for any $1\le j \le L$, $1\le k \le N_{h}$,
where $a_{0}$, $c_{w}$ and $c_{b}$ are complex constants with $|c_{b}|\le |c_{w}|$ to satisfy Condition~\ref{cond:b}, $\tilde{k}=\tilde{k}(k)=\lceil k/L \rceil$, $j_{c}=j_{c}(k)=k-(\tilde{k}-1)L$, $N_{h}$ is an integer multiple of $L$, $\lceil x \rceil$ denotes the ceiling function, and $\tilde{k}_{s}=0$ in this case. It can be shown that such an RBM form can be directly transformed into the RBM form proposed to represent ground states of 1D translationally invariant systems~\cite{carleo2017solving} for any finite $N_{h}$ but we generalize it to an infinitely-many-hidden-node regime ($N_{h}\to\infty$). Since the parameters for different hidden nodes can be generated by the action of a translational-symmetry transformation operator on those for a single hidden node, we just need to focus on one representative hidden node for each level. So we propose an importance measure $\eta(j,\tilde{k},L)$ to measure the importance of a set of edges which is defined as
\begin{IEEEeqnarray}{rCl}
\eta(j,\tilde{k},L) = &&\Big |  \operatorname{Re}(W_{j,(L+1)/2+(\tilde{k}-1)L}^{(L)}) \Big |^{2} \nonumber\\
&+& \beta_{1}^{2} \Big |  \operatorname{Im}(W_{j,(L+1)/2+(\tilde{k}-1)L}^{(L)}) \Big |^{2}
\end{IEEEeqnarray}
and present it as a function of the spin-site index $j$ and level index $\tilde{k}$. Its 3D structure can reflect the decay of both $\lambda(\tilde{k})$ and $\mu(r)$ while the center of the ``orbital'' at every level is localized around $j=(L+1)/2$. So a plotting of the peak at every level as a function of the level index ($\tilde{k}$) can reflect the decay of $\lambda(\tilde{k})$. One example of such LRFD RBM with a power-law decaying $\lambda(\tilde{k})$ is shown in Fig.~\subref*{fig:3Dstandard}.

We show the two types of truncation errors $\varepsilon(L, N_{h})$ as a function of $N_{h}$ with fixed $L$  
for 1D SPT cluster states with a perturbation part (Fig.~\subref*{fig:SPTexp} and \subref*{fig:SPTpower}). It means that the RBM is constructed as a summation of the setting defined in the system of equations~(\ref{eq:SPT}) and a perturbation part specified as Eqs.~(\ref{eq:LRFDW})--(\ref{eq:LRFDa}) show. The numerical results for $\lambda(\tilde{k})$ with exponential and power-law decays are given. As described above, the 1D SPT cluster states can be exactly represented by a short-range ($1$-range) RBM~\cite{deng2017machine}. Using our description, its RBM representation just has one level, and the corresponding $\lambda(\tilde{k})$ and $\mu(r)$ quickly go down to zero for $\tilde{k}> 1$ and $r>1$. The addition of the perturbation part makes the composite RBM a LRFD RBM so that we can study the truncation errors. We give the results for both types of truncation errors and let $\hat{B}$ be the operator of spin correlations between spin $1$ and $2$ in $z$ and $x$ directions. 

Our numerical experiments on the scaling of the truncation errors in $N_{h}$ with fixed $L$ are well upper bounded by our estimations given in inequalities~(\ref{eq:errorPsi}) and (\ref{eq:errorB}), which substantiates our theoretical analysis. Those experiments also indicate that our estimations in Eq.~(\ref{eq:epsvsNh}) correctly capture the asymptotic properties of $\varepsilon(L,N_{h})$ with varying $N_{h}$. Moreover, the fact that the curve of exact $\varepsilon(L,N_{h})$ and that of our estimation associated with $\hat{B}=\hat{\sigma}_{1}^{x}\hat{\sigma}_{2}^{x}$ have exactly the same slope implies that our estimation in Eq.~(\ref{eq:epsvsNh}) gives an asymptotically optimal upper bound. It means that, for the second-type truncation errors (inequality~(\ref{eq:errorB})), there is still room to improve the constant prefactors in our estimation, but we cannot qualitatively further improve the upper bound. In comparison, there is room to both qualitatively improve the upper bound and improve the constant prefactors for the first-type truncation errors (inequality~(\ref{eq:errorPsi})).




\subsection{Scaling of complexity} \label{sec:scaling}

\begin{table*}[tbp]
\caption{\label{table:manifold}
Complexity estimations for distinct typical
settings of $\mu(r)$ and $\lambda(\tilde{k})$.
``$-$'' in the $\mu(r)$ column denotes all $\mu(r)$ functions that make $Q(L)$ converge as $L\to\infty$. ``$-$'' in the $\lambda(\tilde{k})$ column denotes all $\lambda(\tilde{k})$ functions that make $P(N_{h}/L)$ have the asymptotic behavior of $\mathcal{O}( 1/\ln(N_{h}/L) )$ as $N_{h}\to \infty$. Note that $\delta_{P}>1$ and $\alpha_{P}>1/2$ in these settings. }
\begin{ruledtabular}
\begin{tabular}{cccccc}
Manifold&
$\mu(r)$&
$Q(L)$&
$\lambda(\tilde{k})$&
$P(N_{h}/L)$&
$N_{h}^{\ast}(L,\varepsilon)$ \\
\colrule
$S_{2}^{(1)}$ 
& -
& converge
& $\delta_{P}^{-\tilde{k}}$ 
& $\mathcal{O}( \delta_{P}^{-2N_{h}/L} )$ 
& $\mathcal{O}( L\ln (L/\varepsilon) )$ \\
$S_{2}^{(2)}$ 
& -
& converge
& $\tilde{k}^{-\alpha_{P}}$ 
& $\mathcal{O}( (N_{h}/L)^{1-2\alpha_{P}} )$ 
& $\mathcal{O}( (L^{2\alpha_{P}}/\varepsilon)^{1/(2\alpha_{P}-1)} )$ \\
$S_{2}^{(3)}$ 
& $r^{-1}$ $(r\ge 1)$
& $\mathcal{O}( (\ln L)^{2} )$
& $\delta_{P}^{-\tilde{k}}$ 
& $\mathcal{O}( \delta_{P}^{-2N_{h}/L} )$ 
& $\mathcal{O}( L\ln (L/\varepsilon) )$ \\
$S_{2}^{(4)}$ 
& $r^{-1}$ $(r\ge 1)$
& $\mathcal{O}( (\ln L)^{2} )$
& $\tilde{k}^{-\alpha_{P}}$ 
& $\mathcal{O}( (N_{h}/L)^{1-2\alpha_{P}} )$ 
& $\mathcal{O}( (L^{2\alpha_{P}}(\ln L)^{2}/\varepsilon)^{1/(2\alpha_{P}-1)} )$ \\
$S_{2}^{(5)}$ 
& $\to \mu_{\infty}>0$ $(r\to \infty)$
& $\mathcal{O}( L^{2} )$
& $\delta_{P}^{-\tilde{k}}$ 
& $\mathcal{O}( \delta_{P}^{-2N_{h}/L} )$ 
& $\mathcal{O}( L\ln (L/\varepsilon) )$ \\
$S_{2}^{(6)}$ 
& $\to \mu_{\infty}>0$ $(r\to \infty)$
& $\mathcal{O}( L^{2} )$
& $\tilde{k}^{-\alpha_{P}}$ 
& $\mathcal{O}( (N_{h}/L)^{1-2\alpha_{P}} )$ 
& $\mathcal{O}( (L^{2\alpha_{P}+2}/\varepsilon)^{1/(2\alpha_{P}-1)} )$ \\
$S_{2}^{(7)}$ 
& -
& converge
& -
& $\mathcal{O}( 1/\ln(N_{h}/L) )$ 
& $\mathcal{O}(L\exp{(L/\varepsilon)})$\\
\end{tabular}
\end{ruledtabular}
\end{table*}

We can investigate the scaling of spatial complexity in system sizes for LRFD RBMs as the results in Sec.~\ref{sec:LRFD} and Sec.~\ref{sec:trunc} still hold for varying $L$. We give an upper-bound estimation of the complexity of RBM representations which depends on the asymptotic behavior at $x=\infty$ of the functions $P(x)$ (Eq.~(\ref{eq:Px})) and $Q(x)$ (Eq.~(\ref{eq:Qx})), and thus is determined by the decaying rates specified by $\lambda(\tilde{k})$ and $\mu(r)$. 

Define the minimum $N_{h}$ to achieve a sufficiently small approximation error $\varepsilon_{0}$ as
\begin{IEEEeqnarray}{rCl}
N_{h}^{\ast}(L,\varepsilon_{0}) &= &\inf\{ N_{h}:  \varepsilon(L,N_{h}) \le \varepsilon_{0} \}.
\end{IEEEeqnarray}
Using Lemma~\ref{lemma:bound}, the sufficient condition for $\varepsilon(L,N_{h}) \le \varepsilon_{0}$ is that the corresponding upper bound on truncation errors is no larger than $\varepsilon_{0}$. So this provides one way to get an upper bound on $N_{h}^{\ast}(L,\varepsilon_{0})$ for LRFD RBMs. It can be shown that
\begin{IEEEeqnarray}{rCl} \label{eq:NhvsL}
N_{h}^{\ast}(L,\varepsilon_{0}) = \mathcal{O}(L\, p_{d}^{-1}(\frac{L\, q(L)}{f^{-1}(\varepsilon_{0})})) \quad (\textrm{as } L\to \infty), 
\end{IEEEeqnarray}
where $q(x)$, $p_{d}(x)$ and $f(x)$ are functions to specify the asymptotic behaviors of $Q(x)$, $P(x)$ and $F(x)$ as defined above and the superscript ``$^{-1}$'' denotes the inverse of the corresponding function. This upper-bound estimation is usually asymptotically larger than, thus not influenced by, the $n_{\Theta}(L)L$.

Rich information can be extracted from Eq.~(\ref{eq:NhvsL}). First, the first factor $L$ comes from our assumption that $N_{h}$ is an integer multiple of the system size $L$ and the second factor $L$ in front of $q(L)$ is extracted using the translational symmetry of the wave function. So these two factors reflect the growing system sizes and the remaining factors reflect the distinction in complexity for different LRFD RBMs. 

Second, $P(N_{h}/L)$ and $Q(L)$ (thus $\mu(r)$ and $\lambda(\tilde{k})$) which characterize the nonlocal structure of RBMs in our description have qualitatively different influence on the complexity. Specifically, $Q(L)$ can converge to a finite $L$-independent constant in the thermodynamic limit and does not influence the complexity for sufficiently localized ``orbitals'' in the cases where $\mu(r)$ decays sufficiently fast. With the upper boundedness condition for $\mu(r)$ (Eq.~(\ref{eq:mu0})), $Q(L)$ can contribute an at-most-quadratic factor to this upper bound on $N_{h}^{\ast}(L,\varepsilon_{0})$. By contrast, the asymptotic property of $P(x)$ significantly influences the complexity and may lead to the inefficiency of RBM representations if $\lambda(\tilde{k})$ decays sufficiently slowly. That would imply that there are too many high-order correlations between spins to be captured by the RBM so polynomially many parameters are not enough to fully compress the information into the RBM form. But as long as $p_{d}^{-1}(x)$ has an at-most-power-law dependence on $x$, this upper-bound estimation will imply that the complexity is definitely at most polynomial in both system size $L$ and $1/\varepsilon_{0}$ with the above two types of truncation errors. Moreover, it is also remarkable that our estimation only provides an upper bound on the complexity, so a faster-than-polynomial scaling of the bound (such as $S_{2}^{(7)}$ in Table~\ref{table:manifold}) does not necessarily imply the inefficiency of the representation. It is possible that the upper bound is not tight and the real complexity is at most polynomial in this case. 

Third, the asymptotic behavior of $F(x)$ at $x=0$ also influences the scaling of the complexity and it directly acts on $\varepsilon_{0}$. We have demonstrated that, for the two types of truncation errors described above, the corresponding $F(x)$'s ($F_{1}(x)$ and $F_{2}(x)$) are both analytic at $x=0$. For general types of truncation errors that can be upper bounded by a function $F(LQ(L)P(N_{h}/L))$, $1/f^{-1}(\varepsilon_{0})$ has a power-law dependence on $1/\varepsilon_{0}$ as long as $F(x)$ is analytic at $x=0$ based on the Taylor series expansion of the function.

This result suggests separate effects of the factors $\mu(r)$ and $\lambda(\tilde{k})$. The scaling of entanglement entropy, which is an important measure of the complexity of quantum many-body states, is influenced by $\mu(r)$, whereas $\lambda(\tilde{k})$ significantly influences the spatial complexity of parameterization in LRFD RBM representations. The length of the support of $\mu(r)$, which determines the ``range'' $r_{0}$ of RBMs, directly influences the scaling of the entanglement entropy of the states between subregions but does not directly contribute a faster-than-polynomial factor to the parameterization complexity. This result possibly provides further theoretical evidence for the high efficiency of RBMs in representing states with entanglement entropy scaling faster than an area law in system sizes~\cite{deng2017quantum}.    

We apply our complexity estimation to several typical settings of $\mu(r)$ and $\lambda(\tilde{k})$ in Table 
\ref{table:manifold}. The manifolds $S_{2}^{(j)}$ with $1\le j \le 6$ all correspond to a spatial complexity which is at most polynomial in $L$. We also apply this analysis to RBMs constructed as the 1D SPT cluster states with a perturbation part. Our numerical results on the scaling of $N_{h}^{\ast}(L,\varepsilon_{0})$ in $L$ with fixed $\varepsilon_{0}$ (Fig.~\subref*{fig:NhvsL}) for small system sizes are consistent with our theoretical analysis summarized in Table 
\ref{table:manifold}. The piecewise linearity of $N_{h}^{\ast}(L,\varepsilon_{0})$ as a function of $L$ with a slope growing very slowly implies that the scaling is perhaps just slightly faster than linear, consistent with our estimation based on parameter settings. The piecewise linearity is due to our assumption that $N_{h}$ is an integer multiple of $L$. So it applies a ceiling operation to the ratio $N_{h}/L$ which will not change when $L$ varies within a small range. The inset in Fig.~\subref*{fig:NhvsL} shows that $N_{h}^{U}(L,\varepsilon_{0})$ serve as upper bounds on $N_{h}^{\ast}(L,\varepsilon_{0})$ as in our analysis. The $N_{h}^{U}(L,\varepsilon_{0})$ are obtained by using the exact values of the right-hand side of inequality~(\ref{eq:errorPsi}) and its leading-order estimations. These are almost the same and both have a power-law scaling in $L$ as indicated by Eq.~(\ref{eq:NhvsL}), supporting the validity of our complexity analysis.

\subsection{Spin-correlation information} \label{sec:correlation}

\captionsetup[subfigure]{position=top,singlelinecheck=off,justification=raggedright}
\begin{figure}[tbp] 
\centering
{\includegraphics[width=0.38\textwidth] {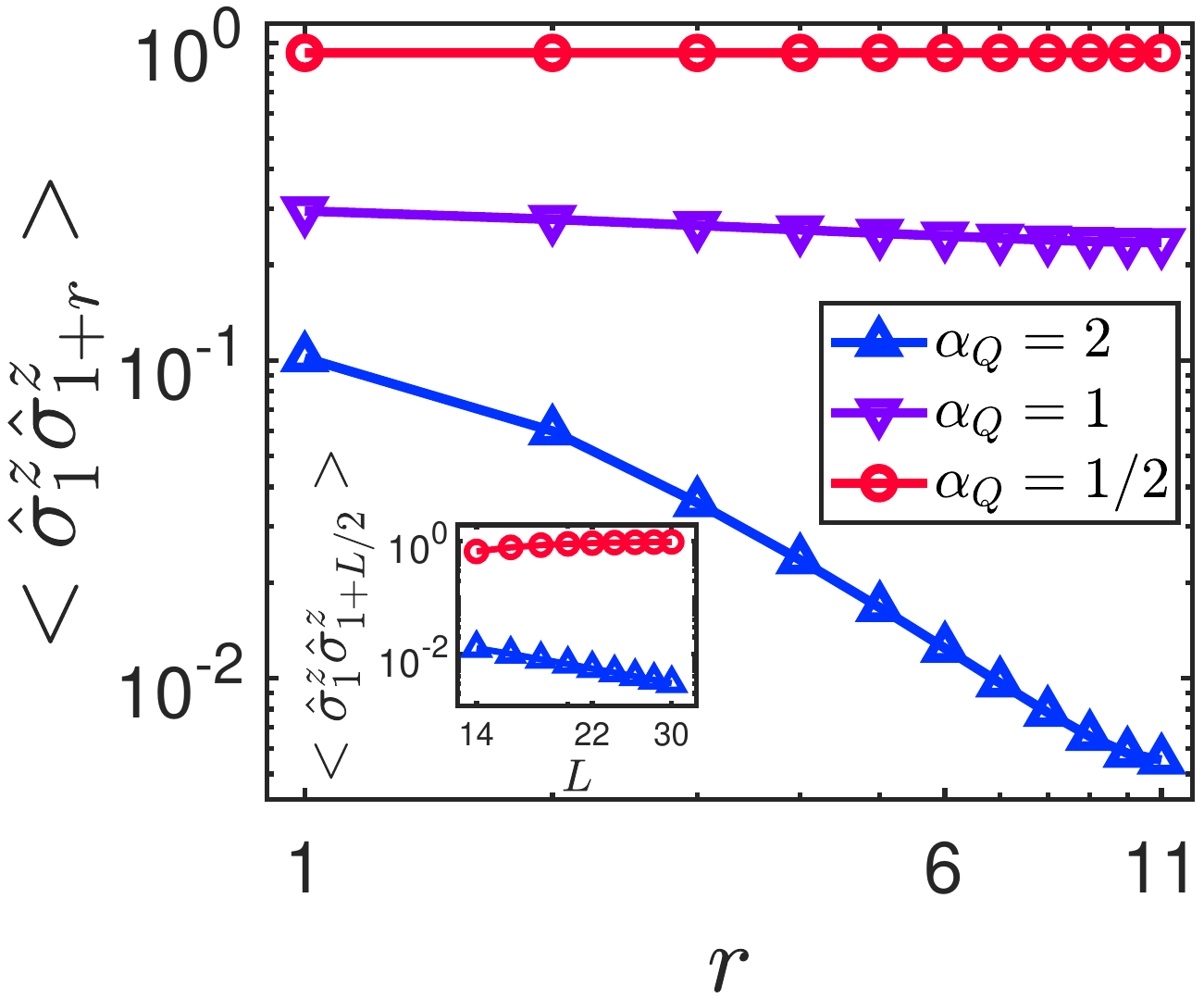} }
\caption{\label{fig:correlation} Spin correlations in the $z$ direction as a function of distance $r$ on a log-log scale. The LRFD RBMs are constructed as Eqs.~(\ref{eq:LRFDW})--(\ref{eq:LRFDa}) show, where $\mu(r)=\frac{1}{2}\delta_{Q}r^{-\alpha_{Q}}$ for $r\neq 0$, $\mu(0)=\delta_{Q}=0.2$, $\lambda(\tilde{k})=\tilde{k}^{-\alpha_{P}}$, $\alpha_{P}=3.5$, $c_{w}=1$, $c_{b}=0$, $a_{0}=0$, $L=22$ and $N_{h}=5L$. The inset shows the spin correlation $\langle \hat{\sigma}_{1}^{z}\hat{\sigma}_{1+L/2}^{z} \rangle$ with $r$ being the half-chain length for varying $L$ (on a log-log scale). It shows a convergence of  $\langle \hat{\sigma}_{1}^{z}\hat{\sigma}_{1+L/2}^{z} \rangle$ to an $L$-independent constant (almost attaining the maximum value of $1$) for $\alpha_{Q}=1/2$ and a decay for $\alpha_{Q}=2$. }
\end{figure}

In this subsection, we analyze what information about the physical properties of the quantum states can be extracted from the LRFD RBM form using our description of the nonlocal structure. Here, we focus on a small-parameter regime in which all $|a_{j}|$, $|b_{k}|$ and $|W_{j,k}|$ are no larger than $\varepsilon_{1}$, and $\varepsilon_{1} \ll 1/L$, $\varepsilon_{1} \ll 1/N_{h}$. We do not explicitly write the superscript ``$^{(L)}$'' for RBM parameters and assume that the RBM just has a finite number ($N_{h}$) of hidden nodes in this subsection.

Based on the proof given in Appendix~\ref{sec:proofcorrelation}, we find that the unnormalized correlation in the $z$ direction between spins with a distance of $r$ for a LRFD RBM with translational symmetry is
\begin{IEEEeqnarray}{rCl}
C_{\textrm{unnorm}}^{z}(r) 
&=& \langle \Psi^{(L,N_{h})} | \hat{\sigma}_{1}^{z}\hat{\sigma}_{1+r}^{z} | \Psi^{(L,N_{h})} \rangle \\
&=& 2\big( \operatorname{Re}(WW^{T}) \big)_{1,1+r} + 4\operatorname{Re}(a_{1})\operatorname{Re}(a_{1+r}) \nonumber\\
&& + \mathcal{O}(\varepsilon_{1}^{3}) \quad (\textrm{as } \varepsilon_{1}\to 0). \label{eq:Crzunnorm}
\end{IEEEeqnarray}
Note that the above result is the $r$-related part of the spin correlation, while the real value of the correlation is $C_{\textrm{unnorm}}^{z}(r)$ divided by an $r$-independent normalization factor $\langle \Psi^{(L,N_{h})} | \Psi^{(L,N_{h})} \rangle$. So for RBMs constructed as Eqs.~(\ref{eq:LRFDW})--(\ref{eq:LRFDa}) show with $a_{0}=0$ and $c_{w}\in \mathbb{R}$ for simplicity,
\begin{IEEEeqnarray}{rCl}
&& C_{\textrm{unnorm}}^{z}(r) \nonumber\\
&\approx& 2|c_{w}|^{2}\sum_{\tilde{k}=1}^{N_{h}/L}|\lambda(\tilde{k})|^{2}\sum_{j_{c}=1}^{L}\mu(|1-j_{\textrm{c}}|_{\textrm{circ}})\mu(|1+r-j_{\textrm{c}}|_{\textrm{circ}}). \nonumber\\
\end{IEEEeqnarray}
So the $\mu(r)$-related factor as shown above describes the decaying rate of spin correlations in the $z$ direction as a function of the distance $r$, while the $\lambda(\tilde{k})$-related factors independent of $r$ do not influence the decaying rate if we only consider the leading-order terms in Eq. (\ref{eq:Crzunnorm}).

The above result in Eq.~(\ref{eq:Crzunnorm}) gives an interpretation of the roles of hidden nodes. The hidden nodes can be viewed as intermediate virtual particles that relate spins (physical particles) at different lattice sites. When an RBM is short-range, the term $\big( \operatorname{Re}(WW^{T}) \big)_{1,1+r}$ will vanish for large enough $r$ as there is no virtual particle that can have both nonzero connectivity to two spins separated by $r$. Then, more intermediate hidden nodes are needed to transport such relations, which means that we need to consider higher-order terms. This is additional evidence that long-range RBMs can represent states with strong quantum correlations. It is shown in Appendix~\ref{sec:proofcorrelation} that, even when $\mu(r)\to 0$ as $r\to 0$, we can still construct LRFD RBMs in which the spin correlations in the $z$ direction can have long-range decayings lower bounded by $\Theta(1/r^{\alpha_{Q}})$ (for $\mu(r)=\Theta(1/r^{\alpha_{Q}})$) with $\alpha_{Q}>1$, $\Theta(\ln r/r)$ (for $\mu(r)=\Theta(1/r)$), and even $\Theta(1)$ (for $\mu(r)=\Theta(1/r^{\alpha_{Q}})$ with $0<\alpha_{Q} \le \frac{1}{2}$). These three kinds of decaying rates of spin correlations are demonstrated by numerical computations (Fig.~\ref{fig:correlation}). The spin correlation $\langle \hat{\sigma}_{1}^{z}\hat{\sigma}_{1+r}^{z} \rangle$ almost saturates the maximum value of $1$ for $\alpha_{Q}=1/2$. In comparison, these spin correlations have different long-range decaying rates for $\alpha_{Q}=1$ and $\alpha_{Q}=2$ as $r$ increases.

\section{Ground-state applications} \label{sec:ground}

\captionsetup[subfigure]{position=top,singlelinecheck=off,justification=raggedright}
\begin{figure}[tbp]
\centering
 \subfloat[]{\includegraphics[width=0.38\textwidth]{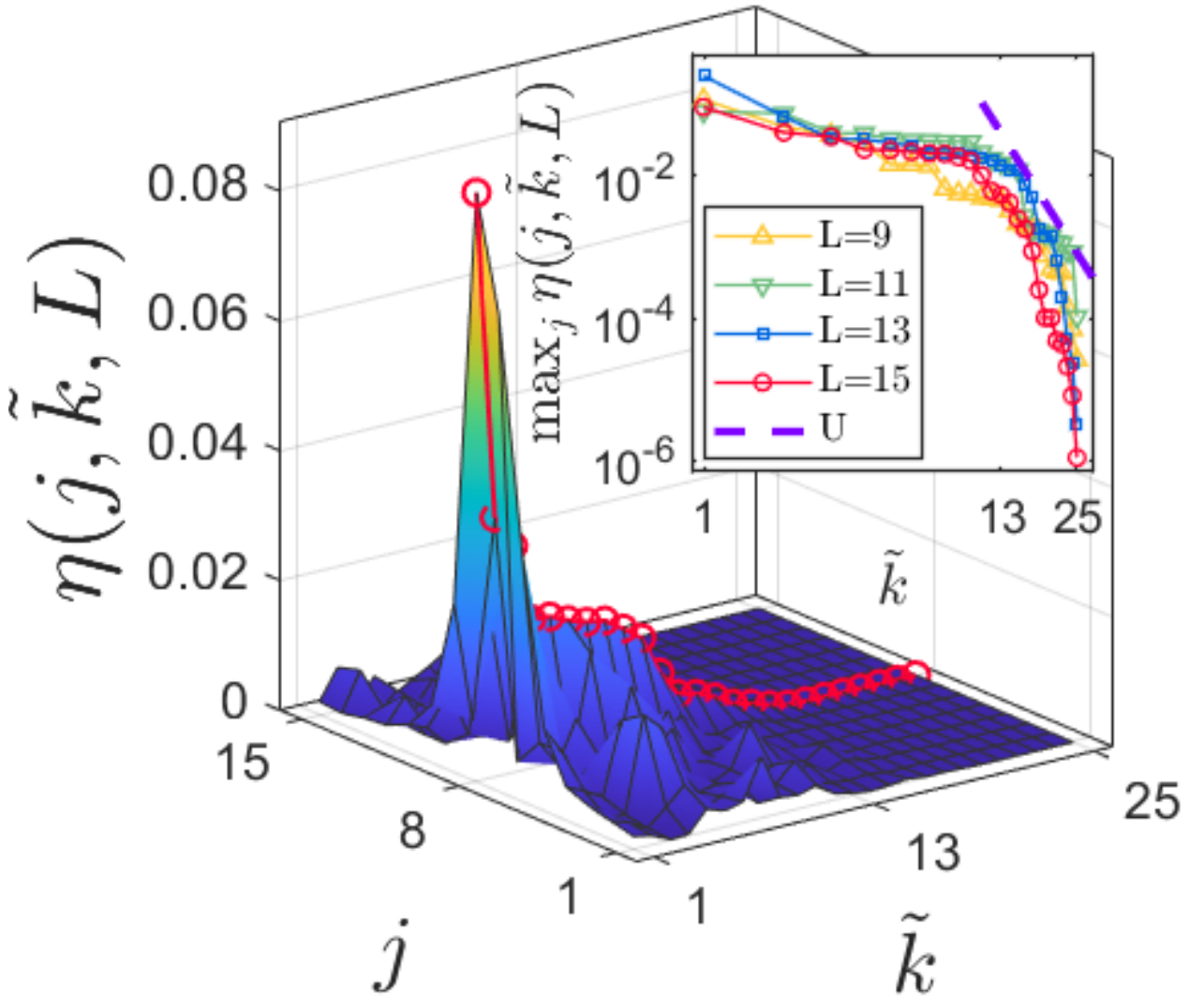} \label{fig:TFIM}} \\
 \subfloat[]{\includegraphics[width=0.38\textwidth]{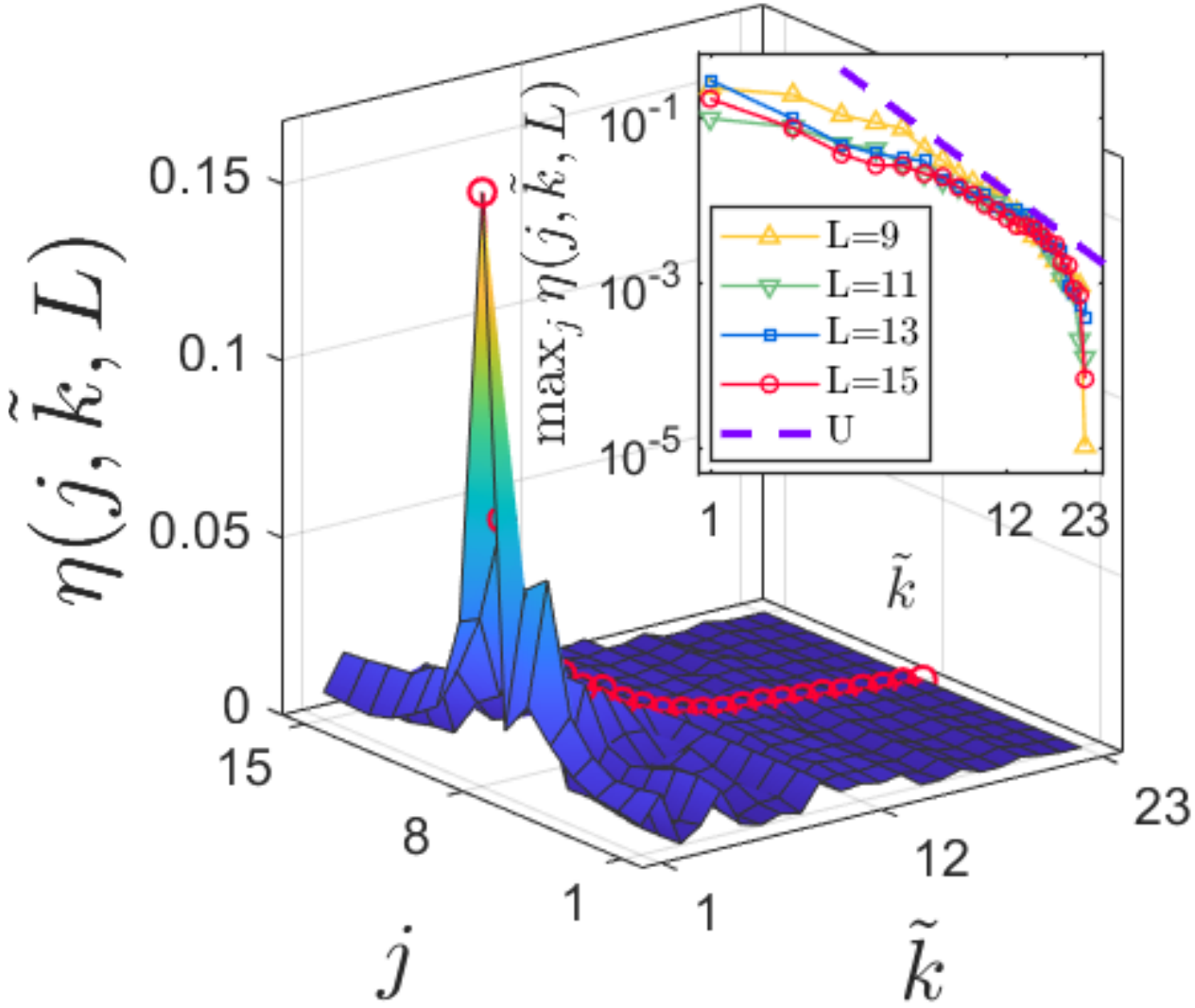} \label{fig:XXZ}} 
 \caption{ Importance measure $\eta(j,\tilde{k},L)$ for the RBMs approximating ground states of two critical systems with $L=15$. (a) TFIM with $B_{x}=1$. (b) XXZ model with $J_{z}=-0.2$. The insets in each subfigure show the decays of the maximum importance measure at each level as level index $\tilde{k}$ increases on a log-log scale. The system size $L=9$, $11$, $13$ and $15$. The purple dashed curve implies that these decaying curves can be upper bounded by a power-law decay. By numerical fitting, the corresponding $\alpha_{P}$ for the dashed lines in the insets of (a) and (b) are $2.957$ and $1.232$, respectively.}
\end{figure}

Based on the proposal of the concept of LRFD RBMs and the theoretical analysis of their spatial complexity, it is natural to explore their applications to learning quantum states associated with specific models. 

First, in Appendix~\ref{sec:spin-up}, we prove that the state with all spins pointing up in the $z$ direction, which is the ground state of a spin-$\frac{1}{2}$ system with a single magnetic field in the $z$ direction and has a form of the Kronecker delta function, can be approximated by LRFD RBMs with arbitrary accuracy. We find that the RBM construction is not unique for such a target state even when fixing the global phase which implies eliminating the degree of freedom associated with a global gauge transformation and we give one construction. Thus, we provide one example of the utility of LRFD RBMs in state representation for arbitrarily large system sizes.

Second, we are particularly interested in the behavior of RBMs in cases where other state representations become less efficient. We numerically study the representation of the ground states of critical systems with finite sizes for which the MPS representation becomes less efficient~\cite{PhysRevB.73.094423,10.5555/2011832.2011833}, while MPS has achieved notable success in representing quantum many-body states with entanglement entropy satisfying an area law~\cite{RevModPhys.93.045003,RevModPhys.82.277}. 

To accomplish this, we use RBMs with translational symmetry and apply the conventional quantum Monte Carlo algorithm (also a variational method) with stochastic-reconfiguration optimizations~\cite{carleo2017solving,sorella2007weak,neuscamman2012optimizing,harju1997stochastic} to learn the ground states of two typical quantum models: the 1D transverse-field Ising model (TFIM) (Eq.~(\ref{eq:TFIM})) and XXZ model (Eq.~(\ref{eq:XXZ})), described by Hamiltonians
\begin{IEEEeqnarray}{rCl}
\hat{H} &=& -\sum_{j=1}^{L} \hat{\sigma}^{z}_{j}\hat{\sigma}^{z}_{j+1}-B_{x}\sum_{j=1}^{L}\hat{\sigma}^{x}_{j}, \label{eq:TFIM}
\end{IEEEeqnarray}
and
\begin{IEEEeqnarray}{rCl}
\hat{H} &=& \sum_{j=1}^{L} (-\hat{\sigma}^{x}_{j}\hat{\sigma}^{x}_{j+1}-\hat{\sigma}^{y}_{j}\hat{\sigma}^{y}_{j+1}+J_{z}\hat{\sigma}^{z}_{j}\hat{\sigma}^{z}_{j+1}) \label{eq:XXZ}
\end{IEEEeqnarray}
with periodic boundary conditions, respectively, where $B_{x}$ denotes the strength of a transverse field and $J_{z}$ denotes the strength of coupling in the $z$ direction. We use RBMs to learn the ground state of the TFIM with $B_{x}=1$ which implies that the quantum system is exactly in the phase-transition point between a ferromagnetic and a paramagnetic phase~\cite{jaschke2017critical} and of the XXZ model with $J_{z}=-0.2$ which implies that the system is in a gapless disordered XY phase~\cite{maghrebi2017continuous}. Both systems are critical systems with the entanglement entropy of the ground states scaling logarithmically in system sizes~\cite{jaschke2017critical,koffel2012entanglement,Chen_2013}. The ground states of these two Hamiltonians (at least for small system sizes) can be well learned by RBMs, which is demonstrated by the high accuracy in spin-correlation calculations given in Appendix~\ref{sec:errorcurve}. The importance measures $\eta(j,\tilde{k},L)$ for these two RBMs are provided in Fig.~\subref*{fig:TFIM} and \subref*{fig:XXZ}.  

The numerical results show that the RBM representations of the two ground states of the above two critical systems have forms very similar to LRFD RBMs. The overall 3D structures for the importance measures $\eta(j,\tilde{k},L)$ are similar to the one presented in Fig.~\subref*{fig:3Dstandard} which corresponds to a standard LRFD RBM. The weight parameters for hidden nodes at the same level are quite localized and decay fastly as the level index $\tilde{k}$ increases and as the spin-site index $j$ goes away from the center. Moreover, it seems that the ``ridge'' of $\eta(j,\tilde{k},L)$ for varying system sizes can be upper bounded by an $L$-independent power-law decay curve, based on which we can extract a corresponding $\alpha_{P}$ characterizing the rate of level decay for these small-system-size wave functions. If these features still hold as $L$ increases and approaches infinity, these states will form LRFD RBMs which belong to the set $S_{2}^{(2)}$ or $S_{2}^{(6)}$ in Table~\ref{table:manifold} and the corresponding $\lambda(\tilde{k})$ and $\mu(r)$ can be defined. 

Moreover, the above results exhibit a feature that is also manifested in the theory of MPS representations. It has been shown that~\cite{PhysRevB.73.094423}, though MPS becomes less efficient in representing the ground states of critical systems, the bond dimension required to achieve an approximation error $\varepsilon_{0}$ can still be upper bounded by a function scaling polynomially in the system size $L$. The exponent in the power-law dependence of spatial complexity of MPSs on $L$ depends on the central charge $c$, which is a quantity
roughly quantifying the ``degrees of freedom of the
theory'' in conformal field theory~\cite{RevModPhys.82.277}. A larger $c$ leads to a higher exponent in that estimation which implies a higher complexity in MPS representation. While the TFIM at the above phase-transition point has $c=\frac{1}{2}$ and the XXZ model in the disordered XY phase has $c=1$~\cite{PhysRevLett.63.708}, our numerical results do show a smaller fitted $\alpha_{P}$ for the XXZ model, which implies that the XXZ model has more intrinsic ``complexity'' compared to the TFIM, thus needing more parameters to capture this complexity.

\section{State manifolds and complexity classification}

Rigorously speaking, the numerical results for systems of finite sizes only provide evidence supporting that the states may be LRFD RBMs but cannot prove it, since the properties of RBMs in the process of approaching the thermodynamic limit are not yet known. Based on the success of RBMs in numerical simulations and the fact that they can often achieve high accuracy even with a constant number of levels (at least for small system sizes), we conjecture that the ground states of a wide range of quantum systems may be exactly represented by LRFD RBMs, or a variant of them. Here, the term ``variant'' means a generalization of the forms specified in Condition~\ref{cond:W} and \ref{cond:b} by including additional factors that can be naturally incorporated into our complexity analysis. For example, the $\lambda(\tilde{k})$ and $\mu(r)$ functional forms, which are $L$-independent in our definition of LRFD RBMs, can be generalized into $\lambda(\tilde{k},L)$ and $\mu(r,L)$, respectively, while their effects can be easily evaluated using our paradigm for complexity analysis.

\captionsetup[subfigure]{position=top,singlelinecheck=off,justification=raggedright}
\begin{figure}[tbp]
\centering
\includegraphics[width=0.4\textwidth]{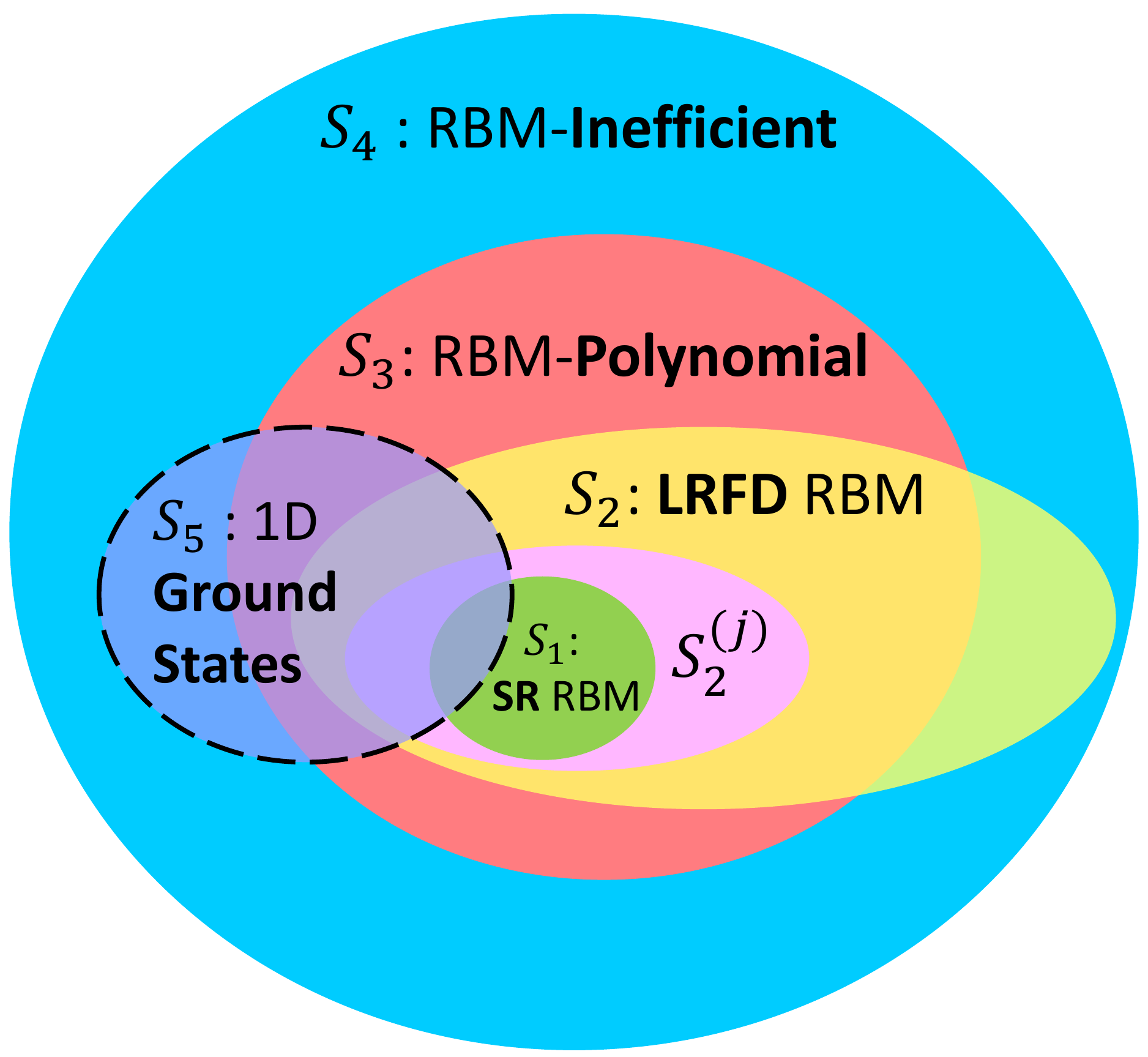}
\caption{\label{fig:manifold} Relations between multiple typical state manifolds. $S_{1}$: short-range RBMs; $S_{2}$: LRFD RBMs; $S_{2}^{(j)}$ (for $1\le j \le 6$, $j\in \mathbb{N}$): LRFD RBMs with distinct parameter conditions, specified in Table~\ref{table:manifold}; $S_{3}$: RBMs with spatial complexities scaling at most polynomially in system sizes; $S_{4}$: RBMs with a faster-than-polynomial scaling of spatial complexities in system sizes, corresponding to inefficiency of representation; $S_{5}$: ground states of 1D quantum spin systems. The dashed boundary of $S_{5}$ means that its relations with other manifolds have not been fully determined.}
\end{figure}

We summarize the relations between multiple typical state manifolds so that the significance of proposing the concept of LRFD RBMs can be better understood. A state manifold usually refers to a subspace of the whole Hilbert space spanned by a parameterized wave-function family~\cite{huang2017complexity}, thus it is a set containing a specific scope of quantum states. The manifolds $S_{1}$, $S_{2}$, $S_{2}^{(j)}$ (for $1\le j \le 6$, $j\in \mathbb{N}$), $S_{3}$ and $S_{4}$ are defined to be the space spanned by quantum states represented by RBMs satisfying corresponding conditions as given in Fig.~\ref{fig:manifold}, while $S_{5}$ is defined to be the manifold spanned by all ground states of 1D quantum many-body spin systems. 

The definitions of these manifolds directly implies that $S_{1} \subsetneq S_{2}^{(j)} \subsetneq S_{2}$ (for $1\le j \le 6$). Our complexity analysis for LRFD RBMs (Sec.~\ref{sec:scaling}) gives the result that $S_{2}^{(j)} \subseteq ( S_{2} \cap S_{3} )$. Previous research shows that a set of problems where RBMs appear to be powerful are related to topological states, among which the 1D SPT cluster states belong to $S_{5} \cap S_{1}$~\cite{deng2017machine}. The Laughlin wave functions, which have the structure of Jastrow wave functions and are associated with chiral topological order, can be exactly represented by RBMs in $S_{3}$ with a quadratic scaling of $N_{h}$ in $L$ but their approximations with RBMs of a long-range form and less complexity are often used~\cite{glasser2018neural}. $S_{4}$ contains all other sets mentioned in Fig.~\ref{fig:manifold} as RBMs without restriction on the number of hidden nodes are universal approximators for discrete distribution~\cite{6796877}. Numerical results seem to support that a ``large fraction'' of $S_{5}$ is contained in its intersection with $S_{2}$. We argue that the concept of $S_{2}$ may benefit the understanding of which fraction of $S_{5}$ falls into its intersection with $S_{3}$, thus also promoting the understanding of the complexity of quantum many-body states.

It is remarkable that our paradigm for complexity analysis and our characterization of the nonlocal structures of RBMs for 1D quantum spin systems can be generalized to higher-dimensional systems, e.g., lattices. This is done by generalizing the description of single-level ``orbitals'' from $\mu(r)$ to $\mu(\vec{r})$ while keeping $\lambda(\tilde{k})$ as a level-decay factor. For deep NN quantum states, we can still view each single hidden layer as a combination of multiple levels which capture correlations of different orders. We can calculate the truncation errors for each hidden layer associated with specific nodal functions and analyze the propagation of errors through layers.

\section{Summary}

In this work, we define a subset of generic RBM quantum states---long-range-fast-decay (LRFD) RBM states. Using the language of levels, the nonlocal structure of LRFD RBMs is described with two functions: one of which, $\mu(r)$, captures the localization of the spatial distribution of the wave function for each single level and encodes information about spin correlations; the other, $\lambda(\tilde{k})$, is a level-decay factor capturing correlations of different orders and significantly influencing the complexity of the RBMs. We derive upper bounds on truncation errors, which allow us to analyze the scaling of the spatial complexity in system sizes and approximation errors for LRFD RBMs. We provide numerical results supporting that the ground states of a wide range of 1D quantum spin systems, including some critical systems, may be approximated by LRFD RBMs with an at-most-polynomial complexity. Finally, we describe the relationships between state manifolds of different computational complexity and identify hierarchies of RBM-efficient approximation. 

Generalizing the RBM wave-function ansatz to an infinitely-many-hidden-node regime and proposing the concept of LRFD RBMs does not imply the use of an infinitely-large neural network for state representations. These serve to define the completeness of a set of variational states and serve as a tool for complexity analysis based on the good extensibility and analyzability of LRFD-RBM forms. This concept may promote general understanding of the intrinsic complexity of quantum many-body states.

\section{Acknowledgments}
We thank Fangli Liu for helpful discussions. This material is based upon work supported by
the U.S. National Science Foundation Physics Frontier Center at the Joint Quantum Institute. The authors acknowledge the University of Maryland supercomputing resources made available for conducting the research reported in this paper. CWC thanks Merton College and the Atomic and Laser Physics sub-department of the University of Oxford for support of part of this work.

\appendix

\section{Proof of the convergence of $|\Psi^{(L,\infty)}\rangle$ for long-range-fast-decay RBMs} \label{sec:proofconverge}

In this section, we prove the convergence of $|\Psi^{(L,\infty)}\rangle$ (Eq.~(\ref{eq:RBMinfty})) which satisfies Conditions~\ref{cond:W} and \ref{cond:b} in the definition of the long-range-fast-decay (LRFD) RBMs given in Sec.~\ref{sec:LRFD}.

We have defined the wave-function amplitude $\psi^{(L,\infty)}(\vec{\sigma})$ for LRFD RBMs
and the amplitude $\psi^{(L,N_{h})}(\vec{\sigma})$ for its corresponding truncated RBMs with the first $N_{h}$ hidden nodes kept.

For any fixed $L$ and any $\vec{\sigma}$, define 
\begin{IEEEeqnarray}{rCl}
A_{n}(\vec{\sigma})=| \psi^{(L,nL)}(\vec{\sigma}) |
\end{IEEEeqnarray}
and 
\begin{IEEEeqnarray}{rCl}
\phi_{n}(\vec{\sigma})=\arg( \psi^{(L,nL)}(\vec{\sigma}) )
\end{IEEEeqnarray}
to be the modulus and the argument of $\psi^{(L,nL)}(\vec{\sigma})$, respectively, as we assume $N_{h}$ to be an integer multiple of $L$ throughout this work. Then
\begin{IEEEeqnarray}{rCl}
\psi^{(L,nL)}(\vec{\sigma})
=A_{n}(\vec{\sigma})e^{i\phi_{n}(\vec{\sigma})}.
\end{IEEEeqnarray}

In the following part, we will prove that the sequence $\{ \psi^{(L,nL)}(\vec{\sigma}): n\in \mathbb{N} \}$ is a Cauchy sequence in $\mathbb{C}$ with $d$ as a metric, where $d: \mathbb{C}\times \mathbb{C}\to \mathbb{R}$ is just the commonly used distance between two complex numbers defined as $d(z_{1},z_{2})=|z_{1}-z_{2}|$ for $z_{1}$, $z_{2}\in\mathbb{C}$.

\begin{proof} 
If there exists any finite $n'\in\mathbb{N}$ such that $\psi^{(L,n'L)}(\vec{\sigma})=0$, then $\psi^{(L,nL)}(\vec{\sigma})=0$ for all $n\ge n'$. Then $\{ \psi^{(L,nL)}(\vec{\sigma}): n\in \mathbb{N} \}$ must be a Cauchy sequence in the metric space $(\mathbb{C},d)$ as it converges to $0$. So we will just focus on the cases where $\psi^{(L,nL)}(\vec{\sigma})\neq 0$, which implies $A_{n}(\vec{\sigma})\neq 0$, for all $n\in\mathbb{N}$ in the following part.

Define the effective angles~\cite{carleo2017solving} $\theta_{k}(\vec{\sigma}) = b_{k}^{(L)}+\sum_{j=1}^{L}\sigma_{j}W_{j,k}^{(L)}$. Then, for any $n > \tilde{k}_{s}, n \in \mathbb{N}$,
\begin{IEEEeqnarray}{rCl}
A_{n}(\vec{\sigma}) &=& \Big| \prod_{j=1}^{L}e^{a_{j}^{(L)}\sigma_{j}}\prod_{k=1}^{nL}\cosh(b_{k}^{(L)}+\sum_{j=1}^{L}\sigma_{j}W_{j,k}^{(L)}) \Big|\\
&=& \Big| \prod_{j=1}^{L}e^{a_{j}^{(L)}\sigma_{j}}\prod_{k=1}^{\tilde{k}_{s}L}\cosh(\theta_{k}(\vec{\sigma})) \Big| \Big| \prod_{k=\tilde{k}_{s}L+1}^{nL}\cosh(\theta_{k}(\vec{\sigma})) \Big| \nonumber\\
\\
&=& A_{0} \Big| \prod_{k=\tilde{k}_{s}L+1}^{nL}\cosh(\theta_{k}(\vec{\sigma})) \Big|,
\end{IEEEeqnarray}
where the contribution of terms with $1\le k\le \tilde{k}_{s}L$ is summarized in an $n$-independent constant $A_{0}=\Big| \prod_{j=1}^{L}e^{a_{j}^{(L)}\sigma_{j}}\prod_{k=1}^{\tilde{k}_{s}L}\cosh(\theta_{k}(\vec{\sigma})) \Big|$ as the boundedness conditions (Conditions~\ref{cond:W} and \ref{cond:b}) only apply to the range $k>\tilde{k}_{s}L$ and the terms in this range will be evaluated separately. 

Define intermediate variables
\begin{IEEEeqnarray}{rCl}
&& u_{k}(\vec{\sigma}) = \operatorname{Re}(b_{k}^{(L)}+\sum_{j=1}^{L}\sigma_{j}W_{j,k}^{(L)}) \in \mathbb{R},\\
&& v_{k}(\vec{\sigma}) = \operatorname{Im}(b_{k}^{(L)}+\sum_{j=1}^{L}\sigma_{j}W_{j,k}^{(L)}) \in \mathbb{R},  \\ 
&& U_{k} = \max_{\{\vec{\sigma}\}}\{|u_{k}(\vec{\sigma})|\}
=|\operatorname{Re}(b_{k}^{(L)})|+\sum_{j=1}^{L}|\operatorname{Re}(W_{j,k}^{(L)})|, \nonumber\\
\\
&& V_{k} = \max_{\{\vec{\sigma}\}}\{|v_{k}(\vec{\sigma})|\}
=|\operatorname{Im}(b_{k}^{(L)})|+\sum_{j=1}^{L}|\operatorname{Im}(W_{j,k}^{(L)})|, \nonumber\\
\end{IEEEeqnarray}
where $U_{k}$ and $V_{k}$ can be viewed as universal measures of the amplitudes of the real and imaginary parts of RBM parameters associated with the $k$-th hidden node regardless of spin configurations. Then
\begin{IEEEeqnarray}{rCl}
A_{n}(\vec{\sigma}) 
= A_{0}\prod_{k=\tilde{k}_{s}L+1}^{nL}\frac{1}{2}\sqrt{ e^{2u_{k}(\vec{\sigma})}+e^{-2u_{k}(\vec{\sigma})}+2\cos(2v_{k}(\vec{\sigma})) }. \nonumber\\
\end{IEEEeqnarray}

So we can upper bound $A_{n}(\vec{\sigma})$ by
\begin{IEEEeqnarray}{rCl}
A_{n}(\vec{\sigma}) 
&\le& A_{0}\prod_{k=\tilde{k}_{s}L+1}^{nL}\cosh(u_{k}(\vec{\sigma})) \\
&\le& A_{0}\prod_{k=\tilde{k}_{s}L+1}^{nL}\exp(\frac{1}{2}u_{k}^{2}(\vec{\sigma})) \\
&\le& A_{0}\prod_{k=\tilde{k}_{s}L+1}^{nL}\exp(\frac{1}{2}U_{k}^{2}) \\
&=& A_{0} \exp \Big( \frac{1}{2}\sum_{k=\tilde{k}_{s}L+1}^{nL} ( |\operatorname{Re}(b_{k}^{(L)})|+\sum_{j=1}^{L}|\operatorname{Re}(W_{j,k}^{(L)})| )^{2} \Big). \nonumber\\
\end{IEEEeqnarray}

Using Conditions~\ref{cond:W} and \ref{cond:b}, we can get an upper bound on $A_{n}(\vec{\sigma})$ expressed in terms of $\lambda_{R}(\tilde{k})$ and $\mu(r)$ as 
\begin{IEEEeqnarray}{rCl}
A_{n}(\vec{\sigma}) 
&\le& A_{0} \exp \Big[ \frac{1}{2}\sum_{k=\tilde{k}_{s}L+1}^{nL} \Big( \lambda_{R}(\tilde{k}(k))\mu(0) \nonumber\\
&& +\sum_{j=1}^{L}\lambda_{R}(\tilde{k}(k)) \mu(|j-j_{\textrm{c}}(k)|_{\textrm{circ}}) \Big)^{2} \Big] \\
&=& A_{0} \exp \Big[ \frac{1}{2}\sum_{k=\tilde{k}_{s}L+1}^{nL} \Big( \lambda_{R}(\tilde{k}(k)) \Big)^{2} \nonumber\\
&& \times \Big( \mu(0) +\sum_{\triangle j=0,\pm 1,...,\pm\frac{L-1}{2}}\mu(|\triangle j|) \Big)^{2} \Big] \\
&=& A_{0} \exp \Big[ 2L \sum_{\tilde{k}=\tilde{k}_{s}+1}^{n} \lambda_{R}^{2}(\tilde{k}) \Big( \sum_{r=0}^{(L-1)/2}\mu(r) \Big)^{2} \Big]. \nonumber\\
\end{IEEEeqnarray}

Note that Eq.~(\ref{eq:P0}) and inequality~(\ref{eq:mu0}) as constraints on $\lambda_{R}(\tilde{k})$ and $\mu(r)$ can be used to upper bound the first and second summations in the exponent of the above expression. Since $\sum_{\tilde{k}=\tilde{k}_{s}+1}^{n} \lambda_{R}^{2}(\tilde{k}) \le \sum_{\tilde{k}=\tilde{k}_{s}+1}^{\infty} \lambda_{R}^{2}(\tilde{k}) \le P_{0}$, we have 
\begin{IEEEeqnarray}{rCl}
A_{n}(\vec{\sigma}) 
\le M_{0} \coloneqq A_{0} \exp \Big( 2LQ(L)P_{0} \Big). \label{eq:M0}
\end{IEEEeqnarray}
Therefore, we prove that, for any fixed $L>0$, $A_{n}(\vec{\sigma})$ is bounded above by an $n$-independent constant $M_{0}$ for all $n > \tilde{k}_{s}$, which is the first step for the overall proof.

Using similar mathematical tricks with a modified range of $k$ and $\tilde{k}$, for any $n > \tilde{k}_{s}$ and $m \in \mathbb{N}$, we can upper bound $| A_{n+m}(\vec{\sigma})/A_{n}(\vec{\sigma}) |$ by
\begin{IEEEeqnarray}{rCl}
\Big| \frac{A_{n+m}(\vec{\sigma})}{A_{n}(\vec{\sigma})} \Big|
&=& \prod_{k=nL+1}^{(n+m)L}\Big| \cosh(\theta_{k}(\vec{\sigma}))  \Big| \\
&\le& \exp \Big[ 2LQ(L) \sum_{\tilde{k}=n+1}^{n+m} \lambda_{R}^{2}(\tilde{k}) \Big].
\end{IEEEeqnarray}

We can also lower bound $| A_{n+m}(\vec{\sigma})/A_{n}(\vec{\sigma}) |$ by
\begin{IEEEeqnarray}{rCl}
\Big| \frac{A_{n+m}(\vec{\sigma})}{A_{n}(\vec{\sigma})} \Big|
&=& \prod_{k=nL+1}^{(n+m)L}\frac{1}{2}\sqrt{ e^{2u_{k}(\vec{\sigma})}+e^{-2u_{k}(\vec{\sigma})}+2\cos(2v_{k}(\vec{\sigma})) } \nonumber\\
\\
&\ge& \prod_{k=nL+1}^{(n+m)L}\frac{1}{2}\sqrt{ 2+2\cos(2v_{k}(\vec{\sigma})) } \\
&=& \prod_{k=nL+1}^{(n+m)L} \Big| \cos(|v_{k}(\vec{\sigma})|) \Big|.
\end{IEEEeqnarray}

Since
\begin{IEEEeqnarray}{rCl}
V_{k} 
\le 2\lambda_{I}(\tilde{k}(k))  \sum_{r=0}^{(L-1)/2}\mu(r),
\end{IEEEeqnarray}
and $\lambda_{I}(\tilde{k}(k))  \to 0$ as $k\to\infty$, we know that
$V_{k}\to 0$ as $k\to\infty$. Define $n_{I} = n_{I}(L) = \min\{ n\ge \tilde{k}_{s}, n\in \mathbb{N}: \forall k\ge n_{I}L+1, V_{k}\le \pi/3 \}$. Using the fact that $|\cos(x)|\ge \exp(-\beta_{1}^{2}x^{2}/2)$ with $\beta_{1}=3\sqrt{2\ln2}/\pi$ for all $x\in [0,\pi/3]$, for $n\ge n_{I}$,
\begin{IEEEeqnarray}{rCl}
\Big| \frac{A_{n+m}(\vec{\sigma})}{A_{n}(\vec{\sigma})} \Big|
&\ge& \prod_{k=nL+1}^{(n+m)L} |\cos(V_{k})| \\
&\ge& \exp(-\frac{1}{2}\beta_{1}^{2}\sum_{k=nL+1}^{(n+m)L}V_{k}^{2}) \\
&\ge& \exp \Big[ -2\beta_{1}^{2}L Q(L) \sum_{\tilde{k}=n+1}^{n+m} \lambda_{I}^{2}(\tilde{k}) \Big]. \nonumber\\
\end{IEEEeqnarray}
In fact, $\cos(x)$ converges to $\exp(-x^{2}/2)$ as $x\to 0$ and a constant $\beta_{1}$ slightly larger than $1$ is introduced to ensure the holding of the inequality while capturing the leading-order asymptotic properties.

Since $P(\tilde{k}_{s})$ converges to a constant $P_{0}/(1+\beta_{1}^{2})$ and all terms are nonnegative, we have $P(m')\to 0$ as $m'\to \infty$. So $\sum_{\tilde{k}=m'+1}^{\infty} \lambda_{R}^{2}(\tilde{k})$ and $\sum_{\tilde{k}=m'+1}^{\infty} \lambda_{I}^{2}(\tilde{k})$ also approach $0$ as $m'\to \infty$. 

Therefore, for any sufficiently small $\varepsilon>0$, there exists $n_{1}\ge n_{I}\ge\tilde{k}_{s}$ such that, for all $m'>n_{1}$, 
\begin{IEEEeqnarray}{rCl}
\sum_{\tilde{k}=m'+1}^{\infty} \lambda_{R}^{2}(\tilde{k}) 
&<& \ln(1+\varepsilon/M_{0})/(2LQ(L)),\\
\sum_{\tilde{k}=m'+1}^{\infty} \lambda_{I}^{2}(\tilde{k}) 
&<& \ln(1/(1-\varepsilon/M_{0}))/(2\beta_{1}^{2}LQ(L)). \nonumber\\
\end{IEEEeqnarray}
Then, for all $n>n_{1}$ and $m\in 
\mathbb{N}$,  
\begin{IEEEeqnarray}{rCl}
\Big| \frac{A_{n+m}(\vec{\sigma})}{A_{n}(\vec{\sigma})} \Big|
&\le& \exp \Big[ 2LQ(L) \sum_{\tilde{k}=n+1}^{\infty} \lambda_{R}^{2}(\tilde{k}) \Big] \\
&<& 1+\varepsilon/M_{0},\\
\Big| \frac{A_{n+m}(\vec{\sigma})}{A_{n}(\vec{\sigma})} \Big|
&\ge& \exp \Big[ -2\beta_{1}^{2}LQ(L) \sum_{\tilde{k}=n+1}^{\infty} \lambda_{I}^{2}(\tilde{k}) \Big] \\
&>& 1-\varepsilon/M_{0},
\end{IEEEeqnarray}
which implies 
\begin{IEEEeqnarray}{rCl}
| A_{n+m}(\vec{\sigma})/A_{n}(\vec{\sigma}) -1| < \varepsilon/M_{0}. \label{eq:Aratio}
\end{IEEEeqnarray}

Therefore, combining inequalities~(\ref{eq:M0}) and (\ref{eq:Aratio}),
\begin{IEEEeqnarray}{rCl}
| A_{n+m}(\vec{\sigma}) - A_{n}(\vec{\sigma})| 
&=& A_{n}(\vec{\sigma}) \Big| \frac{A_{n+m}(\vec{\sigma})}{A_{n}(\vec{\sigma})}-1 \Big| \\
&<& M_{0} \frac{\varepsilon}{M_{0}}=\varepsilon.
\end{IEEEeqnarray}
Thus, we prove that the modulus sequence $\{A_{n}(\vec{\sigma}): n\in \mathbb{N} \}$ is a Cauchy sequence in $\mathbb{R}$. So $A_{n}(\vec{\sigma})$ must converge to some nonnegative constant $A_{\infty}(\vec{\sigma})\in\mathbb{R}$ as $n\to\infty$.

Then, we analyze the argument sequence $\{\phi_{n}(\vec{\sigma}): n\in \mathbb{N} \}$. For $n > \tilde{k}_{s}$ and $m \in \mathbb{N}$, we have
\begin{IEEEeqnarray}{rCl}
&& | \phi_{n+m}(\vec{\sigma})-\phi_{n}(\vec{\sigma}) | \nonumber\\
&=& \Big| \arg\Big( \frac{\psi^{(L,(n+m)L)}(\vec{\sigma})}{ \psi^{(L,nL)}(\vec{\sigma})} \Big) \Big| \\
&=& \Big| \arg \Big( \prod_{k=nL+1}^{(n+m)L} \cosh(u_{k}(\vec{\sigma})) \cos v_{k}(\vec{\sigma}) \nonumber\\
&&+i\sinh(u_{k}(\vec{\sigma})) \sin v_{k}(\vec{\sigma}) \Big) \Big|. 
\end{IEEEeqnarray}
For $n\ge n_{I}>\tilde{k}_{s}$, $|v_{k}(\vec{\sigma})| \le V_{k}\le \pi/3$, so $\cosh(u_{k}(\vec{\sigma})) \cos v_{k}(\vec{\sigma})>0$, which implies that the argument of each term is in the range $(-\pi/2,\pi/2)$. So we have
\begin{IEEEeqnarray}{rCl}
&& | \phi_{n+m}(\vec{\sigma})-\phi_{n}(\vec{\sigma}) | \nonumber\\
&=& \Big| \sum_{k=nL+1}^{(n+m)L} \arctan\Big( \tanh(u_{k}(\vec{\sigma})) \tan(v_{k}(\vec{\sigma})) \Big) \Big| \nonumber\\
\\
&\le& \sum_{k=nL+1}^{(n+m)L}  \arctan\Big( \tanh(| u_{k}(\vec{\sigma}) |) \tan(| v_{k}(\vec{\sigma}) |) \Big).  \nonumber\\
\end{IEEEeqnarray}
Using 
\begin{IEEEeqnarray}{rCl}
\arctan(\tanh(x)\tan(y)) \le \beta_{2}\tanh(x)y \le \beta_{2}xy
\end{IEEEeqnarray}
with $\beta_{2}=3\sqrt{3}/\pi$ for all $x \ge 0$ and $0 \le y \le \pi/3$, we can get
\begin{IEEEeqnarray}{rCl}
&& | \phi_{n+m}(\vec{\sigma})-\phi_{n}(\vec{\sigma}) | \nonumber\\
&\le& \beta_{2}\sum_{k=nL+1}^{(n+m)L}  | u_{k}(\vec{\sigma}) | | v_{k}(\vec{\sigma}) | \\
&\le& \beta_{2}\sum_{k=nL+1}^{(n+m)L}  U_{k}V_{k}\\
&=& \beta_{2}\sum_{k=nL+1}^{(n+m)L} \Big( |\operatorname{Re}(b_{k}^{(L)})|+\sum_{j=1}^{L}|\operatorname{Re}(W_{j,k}^{(L)})| \Big) \nonumber\\
&& \times \Big( |\operatorname{Im}(b_{k}^{(L)})|+\sum_{j=1}^{L}|\operatorname{Im}(W_{j,k}^{(L)})| \Big) \\
&\le& \beta_{2} \Big[ 4L \sum_{\tilde{k}=n+1}^{n+m} \lambda_{R}(\tilde{k})\lambda_{I}(\tilde{k}) \Big( \sum_{r=0}^{(L-1)/2}\mu(r) \Big)^{2} \Big] \\
&=& 4\beta_{2}LQ(L) \sum_{\tilde{k}=n+1}^{n+m} \lambda_{R}(\tilde{k})\lambda_{I}(\tilde{k}).
\end{IEEEeqnarray}

Again, since $\sum_{\tilde{k}=m'+1}^{\infty} \lambda_{R}^{2}(\tilde{k})$ and $\sum_{\tilde{k}=m'+1}^{\infty} \lambda_{I}^{2}(\tilde{k})$ approach $0$ as $m'\to \infty$, $\sum_{\tilde{k}=m'+1}^{\infty} \lambda_{R}(\tilde{k})\lambda_{I}(\tilde{k}) \le \frac{1}{2}\sum_{\tilde{k}=m'+1}^{\infty} (\lambda_{R}^{2}(\tilde{k})+\lambda_{I}^{2}(\tilde{k}) )$ also approaches $0$ as $m'\to \infty$. 

Therefore, for any sufficiently small $\varepsilon>0$, there exists $n_{2}\ge n_{I}\ge\tilde{k}_{s}$ such that, for all $m'>n_{2}$, 
\begin{IEEEeqnarray}{rCl}
\sum_{\tilde{k}=m'+1}^{\infty} \lambda_{R}(\tilde{k})\lambda_{I}(\tilde{k}) < \varepsilon/(4\beta_{2}LQ(L)).
\end{IEEEeqnarray}
Then, for all $n>n_{2}$ and $m\in 
\mathbb{N}$,  
\begin{IEEEeqnarray}{rCl}
&& | \phi_{n+m}(\vec{\sigma})-\phi_{n}(\vec{\sigma}) | \nonumber\\
&<& 4\beta_{2}LQ(L) \frac{\varepsilon}{4\beta_{2}LQ(L)} = \varepsilon.
\end{IEEEeqnarray}
Thus, we prove that the argument sequence $\{\phi_{n}(\vec{\sigma}): n\in \mathbb{N} \}$ is a Cauchy sequence in $\mathbb{R}$. So $\phi_{n}(\vec{\sigma})$ must converge to some constant $\phi_{\infty}(\vec{\sigma})\in\mathbb{R}$ as $n\to\infty$.

Combining the above two conclusions, we can get that, for any sufficiently small $\varepsilon>0$, there exists $n_{3}\ge n_{I}\ge\tilde{k}_{s}$ such that, for all $n>n_{3}$ and $m\in 
\mathbb{N}$, $|A_{n+m}(\vec{\sigma})-A_{n}(\vec{\sigma})|<\varepsilon/\sqrt{2}$ and $|\phi_{n+m}(\vec{\sigma})-\phi_{n}(\vec{\sigma})|<\arccos(1-\frac{\varepsilon^{2}}{4M_{0}^{2}})$. So 
\begin{IEEEeqnarray}{rCl}
&& \Big| \psi^{(L,(n+m)L)}(\vec{\sigma})-\psi^{(L,nL)}(\vec{\sigma}) \Big| \nonumber\\
&=& \Big[ (A_{n+m}(\vec{\sigma})-A_{n}(\vec{\sigma}))^{2} \nonumber\\
&& + 2A_{n+m}(\vec{\sigma})A_{n}(\vec{\sigma})\Big( 1-\cos(\phi_{n+m}(\vec{\sigma})-\phi_{n}(\vec{\sigma})) \Big) \Big]^{1/2} \nonumber\\
\\
&<& \Big[ \frac{1}{2}\varepsilon^{2} + 2M_{0}^{2} \frac{\varepsilon^{2}}{4M_{0}^{2}} \Big]^{1/2} = \varepsilon.
\end{IEEEeqnarray}

Thus, we prove that the sequence $\{ \psi^{(L,nL)}(\vec{\sigma}): n\in \mathbb{N} \}$ is a Cauchy sequence in the metric space $(\mathbb{C},d)$, and $\psi^{(L,nL)}(\vec{\sigma})$ actually converges to the constant $A_{\infty}(\vec{\sigma})e^{i\phi_{\infty}(\vec{\sigma})} \in\mathbb{C}$ as $n\to\infty$.

\end{proof}

\section{Proof of upper bounds on truncation errors for LRFD RBMs} \label{sec:proofbound}

In this section, we provide a proof of the upper bounds on the two types of truncation errors for LRFD RBMs (Lemma~\ref{lemma:bound}) given in Sec.~\ref{sec:trunc}.

We give the proof for the first-type truncation errors as follows.

\begin{proof}
\begin{IEEEeqnarray}{rCl}
&& \Vert |\tilde{\Psi}^{(L,\infty)} \rangle - |\tilde{\Psi}^{(L,N_{h})} \rangle \Vert^{2} \nonumber\\
&=& 2-(\langle \tilde{\Psi}^{(L,\infty)} | \tilde{\Psi}^{(L,N_{h})} \rangle + \textrm{c.c.}) \\
&=& 2-\frac{(\langle \Psi^{(L,\infty)} | \Psi^{(L,N_{h})} \rangle + \textrm{c.c.})}{\sqrt{\langle \Psi^{(L,\infty)} | \Psi^{(L,\infty)} \rangle \langle \Psi^{(L,N_{h})} | \Psi^{(L,N_{h})} \rangle}} \\
&=& 2-\frac{2\sum_{\vec{\sigma}} | \psi^{(L,N_{h})}(\vec{\sigma}) |^{2} \operatorname{Re} \Big( \frac{\psi^{(L,\infty)}(\vec{\sigma})}{\psi^{(L,N_{h})}(\vec{\sigma})} \Big)}{\sqrt{\sum_{\vec{\sigma}} | \psi^{(L,\infty)}(\vec{\sigma}) |^{2} \sum_{\vec{\sigma}} | \psi^{(L,N_{h})}(\vec{\sigma}) |^{2} }}.  
\end{IEEEeqnarray}

Define
\begin{IEEEeqnarray}{rCl}
\Theta = \Theta(L,N_{h}) = \max_{\vec{\sigma}}\{ \Big| \arg\Big( \frac{\psi^{(L,\infty)}(\vec{\sigma})}{ \psi^{(L,N_{h})}(\vec{\sigma})} \Big) \Big| \}
\end{IEEEeqnarray}
to be the maximal argument for the ratio factor $\psi^{(L,\infty)}(\vec{\sigma}) / \psi^{(L,N_{h})}(\vec{\sigma})$.
For $N_{h}/L \ge n_{I}(L)>\tilde{k}_{s}$,
\begin{IEEEeqnarray}{rCl}
&& \Big| \arg\Big( \frac{\psi^{(L,\infty)}(\vec{\sigma})}{ \psi^{(L,N_{h})}(\vec{\sigma})} \Big) \Big| \nonumber\\
&\le& 4\beta_{2}LQ(L) \sum_{\tilde{k}=N_{h}/L+1}^{\infty} \lambda_{R}(\tilde{k})\lambda_{I}(\tilde{k}). 
\end{IEEEeqnarray}
Therefore, 
\begin{IEEEeqnarray}{rCl}
\Theta \le 4\beta_{2}LQ(L) \sum_{\tilde{k}=N_{h}/L+1}^{\infty} \lambda_{R}(\tilde{k})\lambda_{I}(\tilde{k}).
\end{IEEEeqnarray}

Define
\begin{IEEEeqnarray}{rCl}
R_{1} &=& R_{1}(L,N_{h}) = \max_{\vec{\sigma}}\Big\{ \Big| \frac{\psi^{(L,\infty)}(\vec{\sigma})}{ \psi^{(L,N_{h})}(\vec{\sigma})} \Big|^{2} \Big\}
\end{IEEEeqnarray}
and
\begin{IEEEeqnarray}{rCl}
R_{2} &=& R_{2}(L,N_{h}) = \min_{\vec{\sigma}}\Big\{ \Big| \frac{\psi^{(L,\infty)}(\vec{\sigma})}{ \psi^{(L,N_{h})}(\vec{\sigma})} \Big|^{2} \Big\} 
\end{IEEEeqnarray}
to be the maximal and minimal moduli for the square of the ratio factor, respectively. We have
\begin{IEEEeqnarray}{rCl}
R_{1} 
&\le& \exp \Big[ 4LQ(L) \sum_{\tilde{k}=N_{h}/L+1}^{\infty} \lambda_{R}^{2}(\tilde{k}) \Big], \\
R_{2} 
&\ge&  \exp \Big[ -4\beta_{1}^{2}L Q(L) \sum_{\tilde{k}=N_{h}/L+1}^{\infty} \lambda_{I}^{2}(\tilde{k}) \Big].
\end{IEEEeqnarray}

We know that there exists $n_{\Theta}(L) > n_{I}(L) > \tilde{k}_{s}$ such that, for all $N_{h}>n_{\Theta}(L)L$,
\begin{IEEEeqnarray}{rCl}
4\beta_{2}LQ(L) \sum_{\tilde{k}=N_{h}/L+1}^{\infty} \lambda_{R}(\tilde{k})\lambda_{I}(\tilde{k}) \le \frac{\pi}{4}, \label{eq:nTheta}
\end{IEEEeqnarray}
which implies $\Theta \le \pi/4$.
Consider
\begin{IEEEeqnarray}{rCl}
\operatorname{Re} \Big( \frac{\psi^{(L,\infty)}(\vec{\sigma})}{\psi^{(L,N_{h})}(\vec{\sigma})} \Big) 
&\ge& \sqrt{R_{2}}\cos(\Theta),\\
\sum_{\vec{\sigma}} | \psi^{(L,\infty)}(\vec{\sigma}) |^{2} 
&\le& R_{1} \sum_{\vec{\sigma}} | \psi^{(L,N_{h})}(\vec{\sigma}) |^{2}. 
\end{IEEEeqnarray}
We can get
\begin{IEEEeqnarray}{rCl}
&& \Vert |\tilde{\Psi}^{(L,\infty)} \rangle - |\tilde{\Psi}^{(L,N_{h})} \rangle \Vert^{2} \nonumber\\
&\le& 2-\frac{2\sum_{\vec{\sigma}} | \psi^{(L,N_{h})}(\vec{\sigma}) |^{2} \sqrt{R_{2}}\cos(\Theta) }{\sqrt{ R_{1} \Big(  \sum_{\vec{\sigma}} | \psi^{(L,N_{h})}(\vec{\sigma}) |^{2} \Big)^{2} }} \\
&=& 2-2\sqrt{\frac{R_{2}}{R_{1}}}\cos(\Theta) \\
&\le& 2-2 \exp \Big[ -2L Q(L) \sum_{\tilde{k}=N_{h}/L+1}^{\infty} \Big( \lambda_{R}^{2}(\tilde{k})+\beta_{1}^{2}\lambda_{I}^{2}(\tilde{k}) \Big)\Big] \nonumber\\
&& \times \cos \Big[ 4\beta_{2}LQ(L) \sum_{\tilde{k}=N_{h}/L+1}^{\infty} \lambda_{R}(\tilde{k})\lambda_{I}(\tilde{k}) \Big].
\end{IEEEeqnarray}
For simplicity, we have assumed $\lambda_{R}(\tilde{k}) = \lambda_{I}(\tilde{k}) = \lambda(\tilde{k})$ which implies that the real part and imaginary part of RBM parameters have the same decaying rate and 
\begin{IEEEeqnarray}{rCl}
P(N_{h}/L) = \sum_{\tilde{k}=N_{h}/L+1}^{\infty} \lambda^{2}(\tilde{k}).
\end{IEEEeqnarray}
Then
\begin{IEEEeqnarray}{rCl}
&& \Vert |\tilde{\Psi}^{(L,\infty)} \rangle - |\tilde{\Psi}^{(L,N_{h})} \rangle \Vert^{2} \nonumber\\
&\le& 2-2 \exp \Big[ -2(1+\beta_{1}^{2})L Q(L) P(N_{h}/L) \Big] \nonumber\\
&& \times \cos \Big[ 4\beta_{2}LQ(L) P(N_{h}/L) \Big] \\
&=& F_{1}\Big( LQ(L) P(N_{h}/L) \Big),
\end{IEEEeqnarray}
where $F_{1}(x)$ is defined in Eq.~(\ref{eq:F1x}).

The idea of the proof is shown schematically in Fig.~\subref*{fig:ratio}, where $\psi_{\textrm{full}}(\vec{\sigma})$ denotes the amplitude for the full LRFD RBM $\psi^{(L,\infty)}(\vec{\sigma})$ and $\psi_{\textrm{tr}}(\vec{\sigma})$ denotes the amplitude for the truncated RBM $\psi^{(L,N_{h})}(\vec{\sigma})$.
\end{proof}

We give the proof for the second-type truncation errors as follows.

We have defined a Hermitian operator $\hat{B}$ of the form $\hat{B} = \bigotimes_{j=1}^{L} \hat{\sigma}_{j}^{(m_{j})}$ where $\bigotimes$ is the tensor product symbol, $m_{j}\in\{0,1,2,3\}$, $\hat{\sigma}_{j}^{(0)}=I_{2\times 2}$ is the identity matrix, and $\{ \hat{\sigma}_{j}^{(1)}, \hat{\sigma}_{j}^{(2)}, \hat{\sigma}_{j}^{(3)} \}$ denote the Pauli matrices.
\begin{proof}
\begin{IEEEeqnarray}{rCl}
&& | \langle\hat{B}\rangle^{(L,\infty)} - \langle\hat{B}\rangle^{(L,N_{h})} | \nonumber\\
&=& \Big| \frac{ \langle \Psi^{(L,\infty)} | \hat{B} 
|\Psi^{(L,\infty)} \rangle }{ \langle \Psi^{(L,\infty)} | \Psi^{(L,\infty)} \rangle } - \frac{ \langle \Psi^{(L,N_{h})} | \hat{B} 
|\Psi^{(L,N_{h})} \rangle }{ \langle \Psi^{(L,N_{h})} | \Psi^{(L,N_{h})} \rangle } \Big| \nonumber\\
\\ 
&\le& \Big| \frac{ \langle \Psi^{(L,\infty)} | \hat{B} 
|\Psi^{(L,\infty)} \rangle }{ \langle \Psi^{(L,\infty)} | \Psi^{(L,\infty)} \rangle } - \frac{ \langle \Psi^{(L,\infty)} | \hat{B} 
|\Psi^{(L,\infty)} \rangle }{ \langle \Psi^{(L,N_{h})} | \Psi^{(L,N_{h})} \rangle } \Big| \nonumber\\
&& + \Big| \frac{ \langle \Psi^{(L,\infty)} | \hat{B} 
|\Psi^{(L,\infty)} \rangle }{ \langle \Psi^{(L,N_{h})} | \Psi^{(L,N_{h})} \rangle } - \frac{ \langle \Psi^{(L,N_{h})} | \hat{B} 
|\Psi^{(L,N_{h})} \rangle }{ \langle \Psi^{(L,N_{h})} | \Psi^{(L,N_{h})} \rangle } \Big| \nonumber\\
\\
&=& G_{1}+G_{2},
\end{IEEEeqnarray}
where
\begin{IEEEeqnarray}{rCl}
G_{1} 
&=&  \Big| \frac{ \langle \Psi^{(L,\infty)} | \Psi^{(L,\infty)} \rangle }{ \langle \Psi^{(L,N_{h})} | \Psi^{(L,N_{h})} \rangle } - 1 \Big| \Big| \langle\hat{B}\rangle^{(L,\infty)} \Big| 
\end{IEEEeqnarray}
and
\begin{IEEEeqnarray}{rCl}
G_{2} 
&=& \Big| \langle \Psi^{(L,\infty)} | \hat{B} 
|\Psi^{(L,\infty)} \rangle - \langle \Psi^{(L,N_{h})} | \hat{B} |\Psi^{(L,N_{h})} \rangle \Big| \nonumber\\
&& \times ( \langle \Psi^{(L,N_{h})} | \Psi^{(L,N_{h})} \rangle )^{-1} \\
&=& \sum_{\vec{\sigma}_{1}}\sum_{\vec{\sigma}_{2}}  \Big| B_{\vec{\sigma}_{1}\vec{\sigma}_{2}}  \psi^{(L,N_{h})}(\vec{\sigma}_{1})^{*}\psi^{(L,N_{h})}(\vec{\sigma}_{2}) \nonumber\\
&& \times \Big( \frac{\psi^{(L,\infty)}(\vec{\sigma}_{1})^{*}\psi^{(L,\infty)}(\vec{\sigma}_{2})}{\psi^{(L,N_{h})}(\vec{\sigma}_{1})^{*}\psi^{(L,N_{h})}(\vec{\sigma}_{2})} - 1 \Big) \Big| \nonumber\\
&& \times \Big( \sum_{\vec{\sigma}} | \psi^{(L,N_{h})}(\vec{\sigma}) |^{2} \Big)^{-1}
\end{IEEEeqnarray}
capture the contribution of the deviations in the normalization factor and the unnormalized expectation value to the approximation error, respectively, with $B_{\vec{\sigma}_{1}\vec{\sigma}_{2}} = \langle \vec{\sigma}_{1} | \hat{B} |
\vec{\sigma}_{2} \rangle$.

Considering $R_{2} \le \sum_{\vec{\sigma}} | \psi^{(L,\infty)}(\vec{\sigma}) |^{2} / \sum_{\vec{\sigma}} | \psi^{(L,N_{h})}(\vec{\sigma}) |^{2} \le R_{1}$ and $| \langle\hat{B}\rangle^{(L,\infty)} | \le 1$, we can get
\begin{IEEEeqnarray}{rCl}
G_{1} \le \max\{ |R_{1}-1|, |1-R_{2}| \}. 
\end{IEEEeqnarray}
Define
\begin{IEEEeqnarray}{rCl}
\xi = \max_{ (\vec{\sigma}_{1},\vec{\sigma}_{2}) } \Big| \frac{\psi^{(L,\infty)}(\vec{\sigma}_{1})^{*}\psi^{(L,\infty)}(\vec{\sigma}_{2})}{\psi^{(L,N_{h})}(\vec{\sigma}_{1})^{*}\psi^{(L,N_{h})}(\vec{\sigma}_{2})} - 1 \Big|.
\end{IEEEeqnarray}
Then
\begin{IEEEeqnarray}{rCl}
G_{2} 
&\le& \xi \sum_{\vec{\sigma}_{1}}\sum_{\vec{\sigma}_{2}}  \Big| B_{\vec{\sigma}_{1}\vec{\sigma}_{2}} \Big| \Big| \psi^{(L,N_{h})}(\vec{\sigma}_{1})^{*} \Big| \Big| \psi^{(L,N_{h})}(\vec{\sigma}_{2}) \Big| \nonumber\\
&& \times   \Big( \sum_{\vec{\sigma}} | \psi^{(L,N_{h})}(\vec{\sigma}) |^{2} \Big)^{-1} \\
&=& \xi \sum_{\vec{\sigma}_{2}}  \Big| \psi^{(L,N_{h})}(\hat{B}(\vec{\sigma}_{2}))^{*} \Big| \Big| \psi^{(L,N_{h})}(\vec{\sigma}_{2}) \Big| \nonumber\\
&& \times  \Big( \sum_{\vec{\sigma}} | \psi^{(L,N_{h})}(\vec{\sigma}) |^{2} \Big)^{-1}, 
\end{IEEEeqnarray}
where $\hat{B}(\vec{\sigma}_{2})$ is the only spin configuration that makes $B_{\vec{\sigma}_{1}\vec{\sigma}_{2}} \neq 0$ for a specific $\vec{\sigma}_{2}$. Using the Cauchy-Schwarz inequality, 
\begin{IEEEeqnarray}{rCl}
&& \sum_{\vec{\sigma}_{2}}  \Big| \psi^{(L,N_{h})}(\hat{B}(\vec{\sigma}_{2}))^{*} \Big| \Big| \psi^{(L,N_{h})}(\vec{\sigma}_{2}) \Big| \nonumber\\
&\le& \Big( \sum_{\vec{\sigma}_{2}}  \Big| \psi^{(L,N_{h})}(\hat{B}(\vec{\sigma}_{2}))^{*} \Big|^{2} \sum_{\vec{\sigma}_{2}}  \Big| \psi^{(L,N_{h})}(\vec{\sigma}_{2}) \Big|^{2} \Big) ^{1/2}\\
&=& \sum_{\vec{\sigma}} | \psi^{(L,N_{h})}(\vec{\sigma}) |^{2}.
\end{IEEEeqnarray}
Therefore, combined with the geometric features, we can get
\begin{IEEEeqnarray}{rCl}
G_{2} 
&\le& \xi \\ 
&\le& \max \Big\{ \sqrt{(R_{1}\cos(2\Theta)-1)^{2}+R_{1}^{2}\sin^{2}(2\Theta)}, \nonumber\\
&& \sqrt{(R_{2}\cos(2\Theta)-1)^{2}+R_{2}^{2}\sin^{2}(2\Theta)}, \nonumber\\
&& |R_{1}-1|, |1-R_{2}| \Big\} \\
&=& \max \Big\{ \sqrt{R_{1}^{2}-2R_{1}\cos(2\Theta)+1}, \nonumber\\
&& \sqrt{R_{2}^{2}-2R_{2}\cos(2\Theta)+1} \Big\}. 
\end{IEEEeqnarray}
Therefore,
\begin{IEEEeqnarray}{rCl}
&& | \langle\hat{B}\rangle^{(L,\infty)} - \langle\hat{B}\rangle^{(L,N_{h})} | \nonumber\\
&\le& \max\{ |R_{1}-1|, |1-R_{2}| \}  \nonumber\\
&& + \max \Big\{ \sqrt{R_{1}^{2}-2R_{1}\cos(2\Theta)+1}, \nonumber\\
&& \sqrt{R_{2}^{2}-2R_{2}\cos(2\Theta)+1} \Big\} \\
&\le& F_{2}\Big( LQ(L)P(N_{h}/L) \Big), 
\end{IEEEeqnarray}
where $F_{2}(x)$ is defined in Eq.~(\ref{eq:F2x}).

\end{proof}

\section{Proof of spin-correlation formula} \label{sec:proofcorrelation}

In this section, we provide a proof of the relevant spin-correlation formulas in Sec.~\ref{sec:correlation}.

\begin{proof}

We assume that all RBM parameters $a_{j}$'s, $b_{k}$'s and $W_{j,k}$'s are no larger than $\varepsilon_{1}$, and $\varepsilon_{1} \ll 1/L$, $\varepsilon_{1} \ll 1/N_{h}$. In this section, we will not explicitly write the superscript ``$^{(L,N_{h})}$'' and assume that the RBM just has a finite number ($N_{h}$) of hidden nodes. Then we can get the Taylor series expansion of $| \psi(\vec{\sigma}) |^{2} $ about $a_{j}=b_{k}=W_{j,k}=0$ in the small-parameter regime:
\begin{IEEEeqnarray}{rCl}
| \psi(\vec{\sigma}) |^{2}
&=& \exp(\sum_{j=1}^{L}2\operatorname{Re}(a_{j})\sigma_{j}) \prod_{k=1}^{N_{h}}|\cosh(\theta_{k})|^{2} \\
&=& F^{(0)}+\sum_{j_{1}\in\{1,2,\ldots,L\}} F_{j_{1}}^{(1)}\sigma_{j_{1}}+\ldots \\
&=& \sum_{n=0}^{L}\sum_{1\le j_{1} < \cdots < j_{n} \le L} F_{j_{1},\ldots,j_{n}}^{(n)} \prod_{j=j_{1}}^{j_{n}} \sigma_{j},
\end{IEEEeqnarray}
where $F_{j_{1},\ldots,j_{n}}^{(n)}$ is the coefficient associated with the spin array $\prod_{j=j_{1}}^{j_{n}} \sigma_{j}$ only depending on the RBM parameters and the contraction $\sigma_{j}^{2}=1$ has been used.

Then, the unnormalized spin correlation between spin $1$ and spin $1+r$ in the longitudinal ($z$) direction is
\begin{IEEEeqnarray}{rCl}
&& C_{\textrm{unnorm}}^{z}(r) \nonumber\\
&=& \sum_{\sigma_{1}=\pm 1}\cdots \sum_{\sigma_{L}=\pm 1}\sigma_{1}\sigma_{1+r} | \psi(\vec{\sigma}) |^{2} \\
&=& \sum_{\sigma_{1}=\pm 1}\cdots \sum_{\sigma_{L}=\pm 1} \sigma_{1}\sigma_{1+r}  \nonumber\\
&& \times  \sum_{n=0}^{L}\sum_{1\le j_{1} < \cdots < j_{n} \le L} F_{j_{1},\ldots,j_{n}}^{(n)} \prod_{j=j_{1}}^{j_{n}} \sigma_{j}. 
\end{IEEEeqnarray}
By applying the Taylor series expansions to each term, we can get
\begin{IEEEeqnarray}{rCl}
&& \exp(\sum_{j=1}^{L}2\operatorname{Re}(a_{j})\sigma_{j}) \nonumber\\
&=&  \Big( 1+2\sum_{j=1}^{L}\operatorname{Re}(a_{j})^{2} \Big) + 2\sum_{j=1}^{L}\operatorname{Re}(a_{j})\sigma_{j} \nonumber\\
&& +2\sum_{j_{1}=1}^{L}\sum_{1\le j_{2} \le L, j_{2}\neq j_{1}} \operatorname{Re}(a_{j_{1}})\operatorname{Re}(a_{j_{2}})\sigma_{j_{1}}\sigma_{j_{2}} \nonumber\\
&& +\mathcal{O}(\varepsilon_{1}^{3}) \quad (\textrm{as } \varepsilon_{1}\to 0), 
\end{IEEEeqnarray}
\begin{IEEEeqnarray}{rCl}
\theta_{k}^{2}
&=& (b_{k}^{2}+\sum_{j=1}^{L}W_{j,k}^{2}) + 2b_{k}\sum_{j=1}^{L}W_{j,k}\sigma_{j} \nonumber\\
&& + \sum_{j_{1}=1}^{L}\sum_{1\le j_{2} \le L, j_{2}\neq j_{1}} W_{j_{1},k}W_{j_{2},k}\sigma_{j_{1}}\sigma_{j_{2}},
\end{IEEEeqnarray}
and then
\begin{IEEEeqnarray}{rCl}
&& |\cosh(\theta_{k})|^{2} \nonumber\\
&=& 1+\frac{1}{2}\theta_{k}^{2}+\frac{1}{2}(\theta_{k}^{*})^{2}+\mathcal{O}(\varepsilon_{1}^{4}) \\
&=& \Big[ 1+\operatorname{Re}(b_{k}^{2}+\sum_{j=1}^{L}W_{j,k}^{2}) \Big] + 2\operatorname{Re}(b_{k}\sum_{j=1}^{L}W_{j,k})\sigma_{j} \nonumber\\
&& + \operatorname{Re}(\sum_{j_{1}=1}^{L}\sum_{1\le j_{2} \le L, j_{2}\neq j_{1}} W_{j_{1},k}W_{j_{2},k})\sigma_{j_{1}}\sigma_{j_{2}} \nonumber\\
&& +\mathcal{O}(\varepsilon_{1}^{4}) \quad (\textrm{as } \varepsilon_{1}\to 0).
\end{IEEEeqnarray}

Only the $F^{(1)}_{1,1+r}$-related term will not vanish in the summation over all $\sigma_{j}=\pm 1$ as its associated $\prod_{j=j_{1}}^{j_{n}} \sigma_{j}$ exactly cancels $\sigma_{1}\sigma_{1+r}$. Therefore, 
\begin{IEEEeqnarray}{rCl}
C_{\textrm{unnorm}}^{z}(r) 
&=& F^{(1)}_{1,1+r} \\
&=& 2\big( \operatorname{Re}(WW^{T}) \big)_{1,1+r} + 4\operatorname{Re}(a_{1})\operatorname{Re}(a_{1+r}) \nonumber\\
&& + \mathcal{O}(\varepsilon_{1}^{3}) \quad (\textrm{as } \varepsilon_{1}\to 0).
\end{IEEEeqnarray}

So for RBMs constructed as Eqs.~(\ref{eq:LRFDW})--(\ref{eq:LRFDa}) show with $a_{0}=0$ and $c_{w}\in \mathbb{R}$ for simplicity,
\begin{IEEEeqnarray}{rCl}
&& C_{\textrm{unnorm}}^{z}(r) \nonumber\\
&\approx&  2(WW^{T})_{1,1+r} \\
&=& 2|c_{w}|^{2}\sum_{\tilde{k}=1}^{N_{h}/L}|\lambda(\tilde{k})|^{2}\sum_{j_{c}=1}^{L}\mu(|1-j_{\textrm{c}}|_{\textrm{circ}})\mu(|1+r-j_{\textrm{c}}|_{\textrm{circ}}). \nonumber\\
\end{IEEEeqnarray}
Considering the definition of $|\ldots|_{\textrm{circ}}$ and assume $L$ is even without influencing the asymptotic analysis, the $r$-related part in $C_{\textrm{unnorm}}^{z}(r)$ is
\begin{IEEEeqnarray}{rCl}
G^{z}(r) 
&=& \sum_{j_{c}=1}^{L}\mu(|1-j_{\textrm{c}}|_{\textrm{circ}})\mu(|1+r-j_{\textrm{c}}|_{\textrm{circ}})  \\
&=& \sum_{j_{c}=1}^{r-1}\mu(j_{c})\mu(r-j_{c}) +\sum_{j_{c}=L/2+1-r}^{L/2-1}\mu(j_{c})\mu(L-r-j_{c}) \nonumber\\
&& +2\sum_{j_{c}=r+1}^{L/2}\mu(j_{c})\mu(j_{c}-r) + 2\mu(0)\mu(r).
\end{IEEEeqnarray}
Considering the variation of each term with varying $j_{c}$ in the above summation in a ring-shaped geometry corresponding to the periodic boundary conditions, we know that the contribution of the first two summations dominates $G^{z}(r)$ as $r\to L/2$, and the contribution of the second summation is no larger than that from the first one. So we can just focus on the contribution of the first summation, which works as a lower bound on $G^{z}(r)$, when analyzing the asymptotic long-range behavior of $G^{z}(r)$ as $L\to\infty$ and $r\to L/2$.

For $\mu(r)=\delta_{Q}/r$,
\begin{IEEEeqnarray}{rCl}
&& \sum_{j_{c}=1}^{r-1}\mu(j_{c})\mu(r-j_{c}) \nonumber\\
&=& \delta_{Q}^{2}\sum_{j_{c}=1}^{r-1}\frac{1}{r}(\frac{1}{j_{c}}+\frac{1}{r-j_{c}}) = \frac{2\delta_{Q}^{2}}{r}\sum_{j_{c}=1}^{r-1}\frac{1}{j_{c}} \\
&=& 2\delta_{Q}^{2}\frac{\ln r}{r} + \mathcal{O}(\frac{1}{r})  \quad (\textrm{as } L\to \infty, r\to L/2). 
\end{IEEEeqnarray}

For $\mu(r)=\delta_{Q}/r^{\alpha_{Q}}$ with $\alpha_{Q}>1$,
\begin{IEEEeqnarray}{rCl}
&& \sum_{j_{c}=1}^{r-1}\mu(j_{c})\mu(r-j_{c}) \nonumber\\ 
&=& \delta_{Q}^{2}\sum_{j_{c}=1}^{r-1}\frac{1}{r^{\alpha_{Q}}}(\frac{1}{j_{c}}+\frac{1}{r-j_{c}})^{\alpha_{Q}} \\
&\ge& \frac{\delta_{Q}^{2}}{r^{\alpha_{Q}}}\sum_{j_{c}=1}^{r-1}(\frac{1}{j_{c}^{\alpha_{Q}}}+\frac{1}{(r-j_{c})^{\alpha_{Q}}}) 
= \frac{2\delta_{Q}^{2}}{r^{\alpha_{Q}}}\sum_{j_{c}=1}^{r-1}\frac{1}{j_{c}^{\alpha_{Q}}}  \\
&=& 2\delta_{Q}^{2}\zeta(\alpha_{Q})\frac{1}{r^{\alpha_{Q}}} +\mathcal{O}(\frac{1}{r^{2\alpha_{Q}}}) \quad (\textrm{as } L\to \infty, r\to L/2), \nonumber\\
\end{IEEEeqnarray}
where $\zeta(x)$ denotes the Riemann zeta function~\cite{titchmarsh1986theory}.

For $\mu(r)=\delta_{Q}/r^{\alpha_{Q}}$ with $0<\alpha_{Q}\le \frac{1}{2}$,
\begin{IEEEeqnarray}{rCl}
G^{z}(\frac{L}{2}) 
&=& 2\sum_{j_{c}=1}^{\frac{L}{2}-1}\mu(j_{c})\mu(\frac{L}{2}-j_{c})+2\mu(0)\mu(\frac{L}{2}) \\
&\ge& 2\delta_{Q}^{2}\sum_{j_{c}=1}^{L/2-1}\frac{1}{j_{c}^{\alpha_{Q}}}\frac{1}{(L/2-j_{c})^{\alpha_{Q}}}  \\
&\ge& 2\delta_{Q}^{2}\sum_{j_{c}=1}^{L/2-1}\frac{1}{[ (j_{c}+L/2-j_{c})/2]^{2\alpha_{Q}}}  \\
&\ge& 2\delta_{Q}^{2} \frac{L/2-1}{L/4} \\
&=& 4\delta_{Q}^{2}+o(1) \quad (\textrm{as } L\to \infty, r\to L/2). \nonumber\\
\end{IEEEeqnarray}

\end{proof}

While the truncation error for a geometric series can be directly estimated, we derive an estimation of the order of the truncation error for a series with terms having a power-law-decaying form as follows. This result serves for the complexity analysis in this work. We know that~\cite{titchmarsh1986theory}, for any $s>1$, $b \ge a\ge 1$, and $a,b\in \mathbb{N}$,
\begin{IEEEeqnarray}{rCl}
\sum_{n=a+1}^{b}\frac{1}{n^{s}} 
&=& \frac{a^{1-s} -b^{1-s}}{s-1} -s\int_{a}^{b}\frac{(x-\lfloor x \rfloor -\frac{1}{2})}{x^{s+1}}\textrm{d}x \nonumber\\
&& +\frac{b^{-s}-a^{-s}}{2}, 
\end{IEEEeqnarray}
where $\lfloor x \rfloor$ denotes the floor function. Consider
\begin{IEEEeqnarray}{rCl}
| s\int_{a}^{b}\frac{(x-\lfloor x \rfloor -\frac{1}{2})}{x^{s+1}}\textrm{d}x| 
&\le& \frac{s}{2}\int_{a}^{b}\frac{1} {x^{s+1}}\textrm{d}x \\
&=& \frac{a^{-s}-b^{-s}}{2}.
\end{IEEEeqnarray}
Therefore,
\begin{IEEEeqnarray}{rCl}
\sum_{n=a+1}^{b}\frac{1}{n^{s}} = \frac{a^{1-s} -b^{1-s}}{s-1} + \mathcal{O}(a^{-s}) \quad (\textrm{as } a\to \infty). \nonumber\\
\end{IEEEeqnarray}
The above results also hold for $b\to\infty$. Therefore,
\begin{IEEEeqnarray}{rCl}
\sum_{n=a+1}^{\infty}\frac{1}{n^{s}} = \frac{a^{1-s}}{s-1} + \mathcal{O}(a^{-s}) = \mathcal{O}(a^{1-s}) \quad (\textrm{as } a\to \infty). \nonumber\\
\end{IEEEeqnarray}

\section{LRFD RBMs approximating the Kronecker delta function} \label{sec:spin-up}

\captionsetup[subfigure]{position=top,singlelinecheck=off,justification=raggedright}
\begin{figure}[tbp]
\centering
\includegraphics[width=0.4\textwidth]{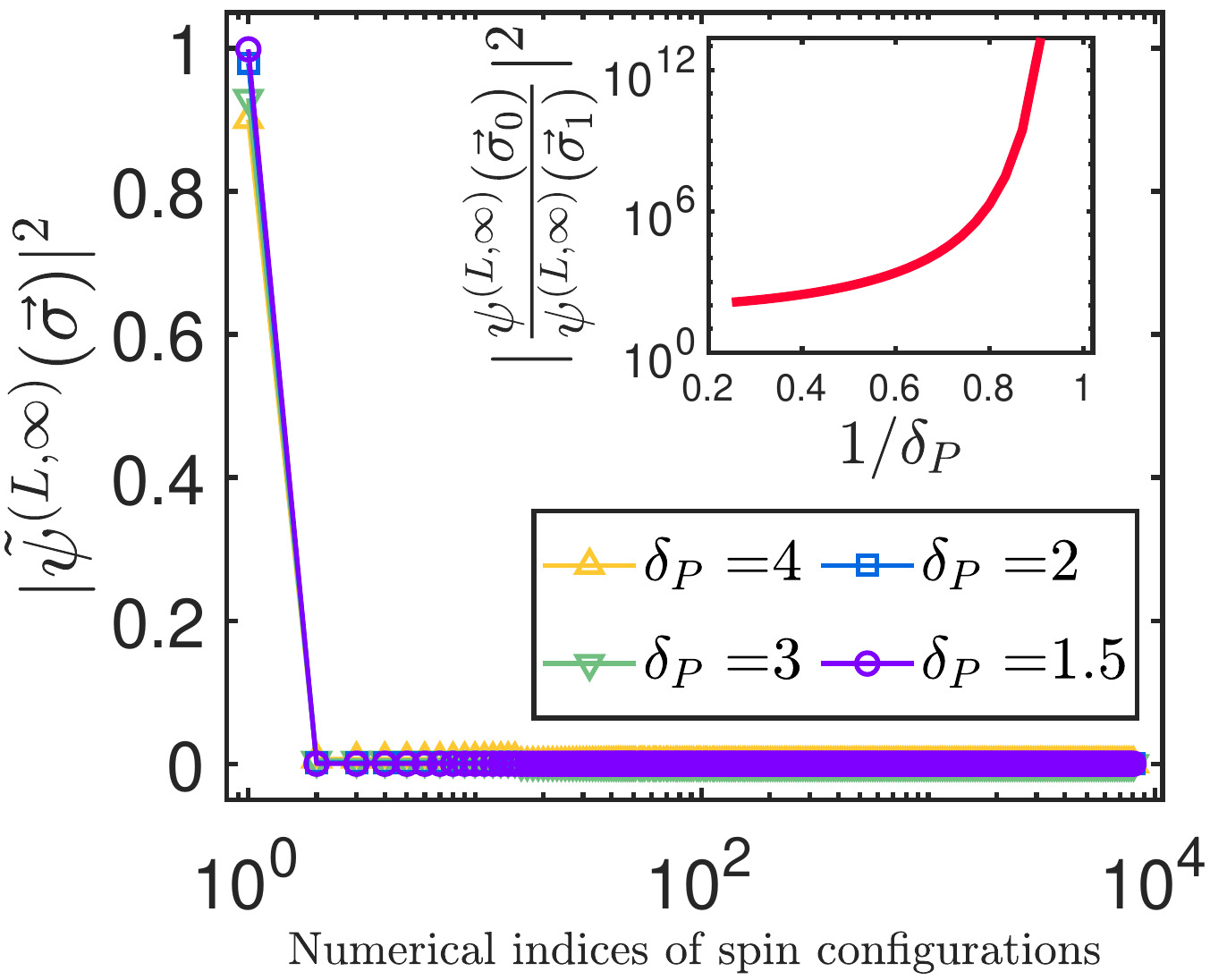}
\caption{\label{fig:Krondelta} Distribution of the square of normalized wave-function amplitudes in the spin-configuration space which can approximate the Kronecker delta function. The LRFD RBMs are constructed as Eqs.~(\ref{eq:Kronmu})--(\ref{eq:Krona}) show and $\mu_{0}=0.1$, $L=13$ and $\lambda(\tilde{k})=\delta_{P}^{1-\tilde{k}}$ with $\delta_{P}>1$. The horizontal axis denotes the numerical indices of all $2^L$ spin configurations which are sorted in a monotonically decreasing order of their corresponding amplitudes. The inset shows the ratio $| \psi^{(L,\infty)}(\vec{\sigma}_{0}) / \psi^{(L,\infty)}(\vec{\sigma}_{1})  |^{2}$ as a function of $1/\delta_{P}$.}
\end{figure}

In this section, we construct LRFD RBMs which can approximate the quantum state with all spins pointing up in the $z$ direction, which has a form of the Kronecker delta function, with arbitrary accuracy for any finite system size. The RBMs are constructed as Eqs.~(\ref{eq:LRFDW})--(\ref{eq:LRFDa}) show with
\begin{IEEEeqnarray}{rCl}
\mu(r) &=& \mu_{0} >0 \quad (\textrm{for any } r\ge 0), \label{eq:Kronmu}\\
c_{w} &=& c_{b}=1, \label{eq:Kronc}\\
a_{0} &=& 0. \label{eq:Krona}
\end{IEEEeqnarray}

Since all $b_{k}$ and $W_{j,k}$ parameters for these RBMs are real positive numbers, the wave-function amplitude for spin configurations reaches its maximum at $\vec{\sigma}_{0} = (1,1,\ldots,1)$ which corresponds to all spins up. We can get that, for any other spin configurations $\vec{\sigma}'$,
\begin{IEEEeqnarray}{rCl}
&& | \psi^{(L,\infty)}(\vec{\sigma}_{0}) / \psi^{(L,\infty)}(\vec{\sigma}')  | \nonumber\\
&\ge& | \psi^{(L,\infty)}(\vec{\sigma}_{0}) / \psi^{(L,\infty)}(\vec{\sigma}_{1})  | \\
&=& \prod_{\tilde{k}=1}^{\infty} \Big[ \frac{ \cosh((L+1)\mu_{0}\lambda(\tilde{k})) }{ \cosh((L-1)\mu_{0}\lambda(\tilde{k})) } \Big]^{L},
\end{IEEEeqnarray}
where $\vec{\sigma}_{1} = (1,1,\ldots,1,-1)$ refers to the spin configuration obtained by flipping the last spin in $\vec{\sigma}_{0}$. Let $x=(L-1)\mu_{0}\lambda(\tilde{k})$ and $\triangle x=2\mu_{0}\lambda(\tilde{k})$. By performing the Taylor series expansion of $\cosh(x+\triangle x)$ about $x$ and comparing the leading-order terms, we can get that there exist constants $0<\beta_{3}<1$ and $x_{0}>0$ such that, for any $0< \triangle x < x < x_{0}$,
\begin{IEEEeqnarray}{rCl}
\frac{\cosh(x+\triangle x)}{\cosh(x)} 
\ge 1+\tanh(x)\triangle x \ge e^{\beta_{3}x\triangle x}.
\end{IEEEeqnarray}
For any fixed $L$ and $\mu_{0}$, there exists $\tilde{k}_{0}\in \mathbb{N}$ such that $(L-1)\mu_{0}\lambda(\tilde{k}) < x_{0}$ for all $\tilde{k}>\tilde{k}_{0}$. So it can be proven that 
\begin{IEEEeqnarray}{rCl}
&& | \psi^{(L,\infty)}(\vec{\sigma}_{0}) / \psi^{(L,\infty)}(\vec{\sigma}')  | \nonumber\\
&\ge& \prod_{\tilde{k}=1}^{\tilde{k}_{0}} \Big[ \frac{ \cosh((L+1)\mu_{0}\lambda(\tilde{k})) }{ \cosh((L-1)\mu_{0}\lambda(\tilde{k})) } \Big]^{L} \nonumber\\
&& \times \prod_{\tilde{k}=\tilde{k}_{0}+1}^{\infty} \Big[ \exp(2\beta_{3}(L-1)\mu_{0}^{2}\lambda^{2}(\tilde{k}))  \Big]^{L} \\
&\ge&  \exp \big( 2\beta_{3}L(L-1)\mu_{0}^{2}\sum_{\tilde{k}=\tilde{k}_{0}+1}^{\infty}\lambda^{2}(\tilde{k}) \big). \label{eq:Kronlr}
\end{IEEEeqnarray}
The above lower bound contains infinitely many terms, though each of which is upper bounded by an expression associated with $x_{0}$, and can reach arbitrarily high values by decreasing the decaying rate of $\lambda(\tilde{k})$. Therefore, the long-range nature of these RBMs allows the shape of $\psi^{(L,\infty)}(\vec{\sigma})$ in the spin-configuration space to approach the Kronecker delta function as $\lambda(\tilde{k})$ decays more and more slowly. Our numerical results in Fig.~\ref{fig:Krondelta} with $\lambda(\tilde{k})=\delta_{P}^{-(\tilde{k}-1)}$ imply that the distribution of the square of normalized wave-function amplitudes in the spin-configuration space can approximate the Kronecker delta function with increasing accuracy as $1/\delta_{P}$ grows up and the ratio $| \psi^{(L,\infty)}(\vec{\sigma}_{0}) / \psi^{(L,\infty)}(\vec{\sigma}_{1})  |^{2}$ as a measure of the approximation accuracy can reach arbitrarily high values as $1/\delta_{P}$ approaches $1$, thus supporting our argument.

\section{Error curves} \label{sec:errorcurve}

\captionsetup[subfigure]{position=top,singlelinecheck=off,justification=raggedright}
\begin{figure}[tbp]
\centering
 \subfloat[]{\includegraphics[width=0.25\textwidth]{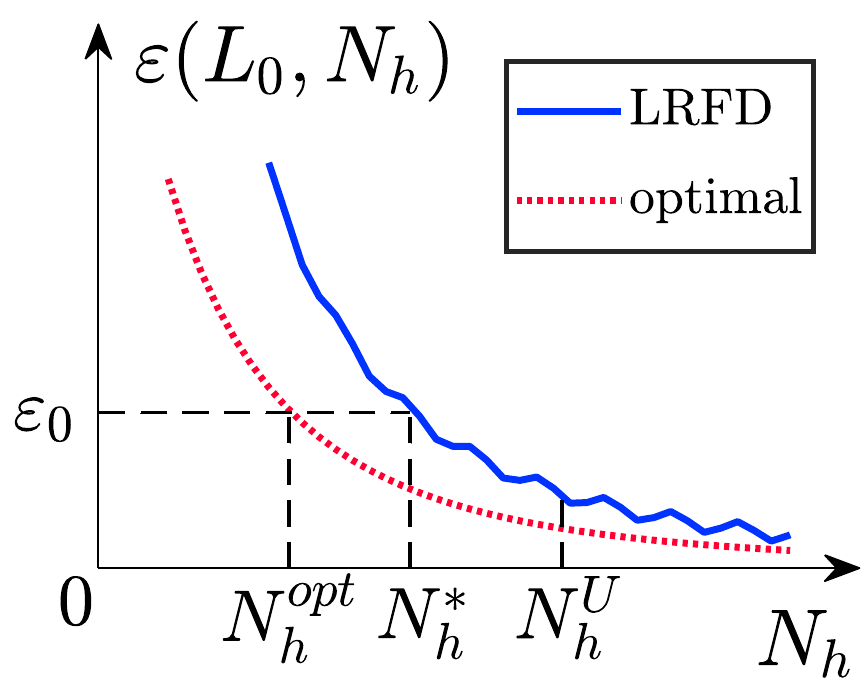} \label{fig:erropt}} 
 \subfloat[]{\includegraphics[width=0.25\textwidth]{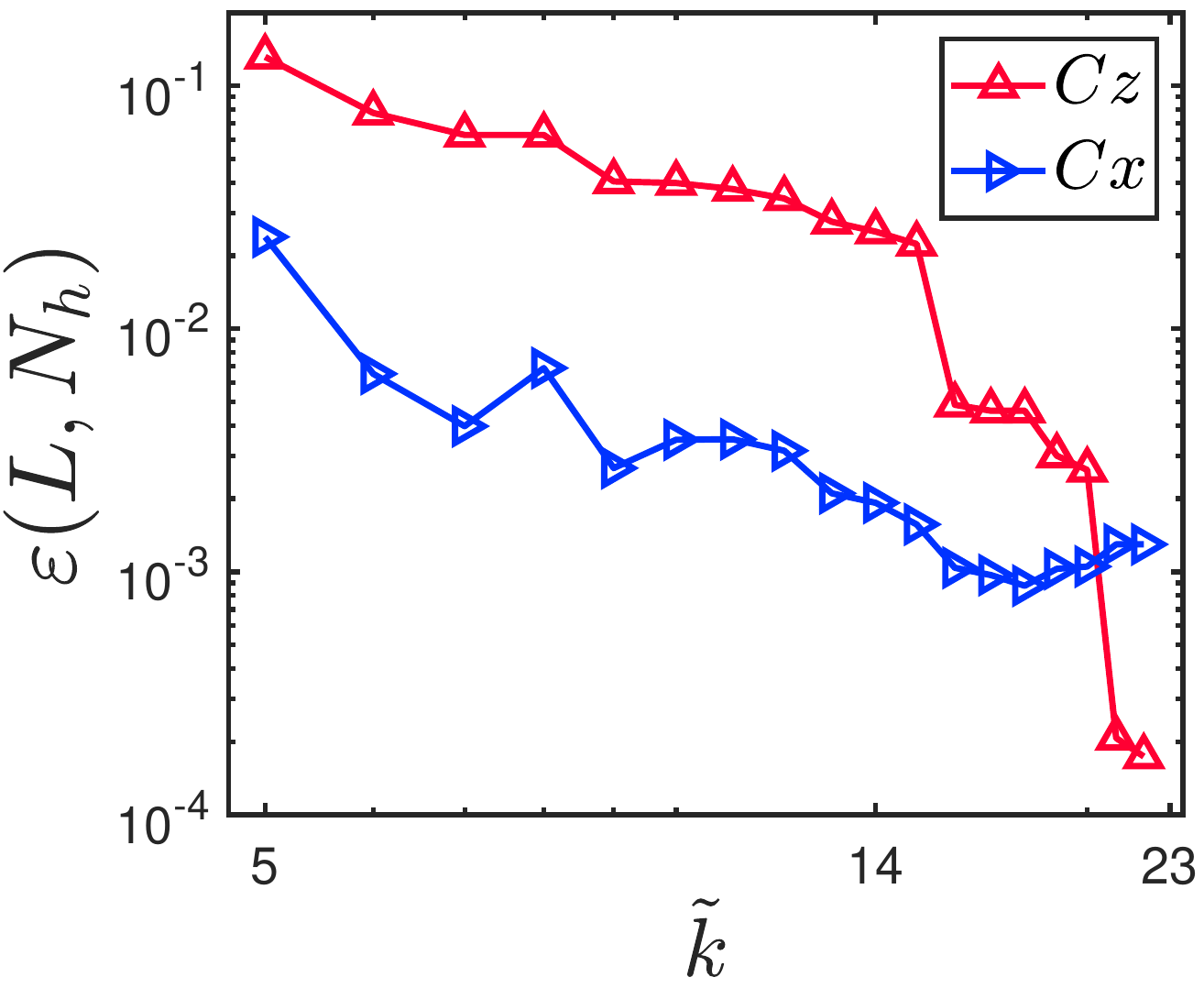} \label{fig:CxCz}} 
 \caption{(a) Approximation errors as a function of the number of hidden nodes for truncated LRFD RBMs and the optimal RBMs. (b) Approximation errors for the calculations of spin correlations in the $x$ and $z$ directions as a function of the number of levels kept in the truncated LRFD RBMs for the XXZ model compared with results from exact-diagonalization methods. The figure is plotted on a log-log scale.}
\end{figure}

In this section, we provide the approximation errors as a function of the number of hidden nodes for truncated LRFD RBMs (Fig.~\subref*{fig:erropt}) and the optimal RBMs as well as the truncation errors for the calculations of spin correlations in the $x$ and $z$ directions as a function of the number of levels kept in the truncated LRFD RBMs (Fig.~\subref*{fig:CxCz}). In Fig.~\subref*{fig:erropt}, $\varepsilon(L_{0},N_{h})$ for LRFD RBMs denotes the second-type truncation errors with $\hat{B}$ being generalized into the Hamiltonian of the quantum system in the ground-state learning which usually has a form of a linear combination of polynomially many original $\hat{B}$-type operators. Then the approximation accuracy of using the optimal RBM with the number of hidden nodes not exceeding $N_{h}$ is definitely better (at least no worse) than that of using an RBM which is a finite truncation of a LRFD RBM keeping $N_{h}$ hidden nodes based on definitions. So $N_{h}^{U}(L_{0},\varepsilon_{0})$ in our complexity analysis actually also provides an upper bound on $N_{h}^{\textrm{opt}}(L_{0},\varepsilon_{0})$ defined as the minimum number of hidden nodes to achieve a specific approximation error $\varepsilon_{0}$ for any kinds of RBMs (not limited to LRFD RBMs), which is of great importance for pre-training computational-resource estimations in ML tasks. The truncation error for LRFD RBMs will converge to $0$ and is possibly not monotonically decreasing as $N_{h}$ approaches infinity. Fig.~\subref*{fig:CxCz} shows that the decaying curves of the second-type truncation errors for the ground-state learning of the XXZ model with the same parameter setting as in Fig.~\subref*{fig:XXZ} are consistent with our analysis and can be upper bounded by power-law decaying curves and reach high accuracy as the number of preserved levels increases.


%

\end{document}